\documentclass[longauth]{aa}
\usepackage[utf8]{inputenc}

\usepackage{graphicx}
\usepackage{natbib}
\usepackage{txfonts}
\usepackage{float}

\newcommand\hyp{{\em HYPERION}}
\newcommand\hyph{{\em XMM-HYPERION}}
\newcommand\xmm{{\em XMM-Newton}}
\newcommand\chandra{{\em Chandra}}

\newcommand\kms{\rm{km\,s^{-1}}}
\newcommand{\fluxcgs}{\rm erg~s^{-1}~cm^{-2}}
\newcommand{\lumcgs}{\rm erg~s^{-1}}

\newcommand{\nhi}{\rm cm^{-2}}
\newcommand{\nhsym}{N_{\rm H}}

\newcommand{\lumh}{L_{\rm 2-10}}
\newcommand{\luvmon}{L_{\rm 2500\rm\AA}}
\newcommand{\lxmon}{L_{\rm 2\rm~keV}}

\newcommand{\msun}{M_{\rm \odot}}
\newcommand{\lsun}{L_{\odot}}

\newcommand{\lbol}{L_{\scriptstyle \rm bol}}
\newcommand{\kbol}{K_{\scriptstyle \rm bol}^{\scriptstyle \rm X}}
\newcommand{\ledd}{\lambda_{\rm Edd}}
\newcommand{\lumedd}{L_{\rm Edd}}

\newcommand{\aox}{\alpha_{\scriptscriptstyle \rm OX}}
\newcommand{\mbh}{M_{\scriptscriptstyle \rm BH}}
\newcommand{\mseed}{M_{\scriptscriptstyle BH}^{\scriptscriptstyle seed}}

\begin{document}

\title{HYPerluminous quasars at the Epoch of ReionizatION (\hyp). A new regime for the X-ray nuclear properties of the first quasars }
   \titlerunning{Hyperluminous QSO at Reionization Epoch. X-ray properties}
   
   \author{L. Zappacosta\inst{1} \and 
          E. Piconcelli\inst{1} \and
          F. Fiore\inst{2,3} \and
          I. Saccheo\inst{1,4} \and
          R. Valiante\inst{1,5} \and
          C. Vignali\inst{6,7} \and
          F. Vito\inst{7} \and
          M. Volonteri\inst{8} \and
          M. Bischetti\inst{2,9} \and
          A. Comastri\inst{7} \and
          C. Done\inst{10}  \and
          M. Elvis\inst{11} \and
          E. Giallongo\inst{1} \and
          F. La Franca\inst{1,4} \and
          G. Lanzuisi\inst{7} \and
          M. Laurenti\inst{12,13} \and
          G. Miniutti\inst{14} \and
          A. Bongiorno\inst{1} \and
          M. Brusa\inst{6,7} \and
          F. Civano\inst{15} \and
          S. Carniani\inst{16} \and
          V. D'Odorico\inst{2,16,3} \and
          C. Feruglio\inst{2,3} \and
          S. Gallerani\inst{16} \and
          R. Gilli\inst{7} \and
          A. Grazian\inst{17} \and
          M. Guainazzi\inst{18} \and
          A. Marinucci\inst{19} \and
          N. Menci\inst{1} \and
          R. Middei\inst{1,13} \and
          F. Nicastro\inst{1} \and
          S. Puccetti\inst{19} \and
          F. Tombesi\inst{1,15,20,21} \and
          A. Tortosa\inst{1} \and
          V. Testa\inst{1} \and
          G. Vietri\inst{22} \and
          S. Cristiani\inst{2,3,23} \and
          F. Haardt\inst{24,25,26} \and
          R. Maiolino\inst{27,28,29} \and
          R. Schneider\inst{1,30,5,31} \and
          R. Tripodi\inst{2,3,9} \and
          L. Vallini\inst{7} \and
          E. Vanzella\inst{7}
          }

   \institute{INAF - Osservatorio Astronomico di Roma, via di Frascati 33, 00078 Monte Porzio Catone, Italy \and
   INAF - Osservatorio Astronomico di Trieste, Via G. Tiepolo 11, I-34143 Trieste, Italy \and
   IFPU - Institute for Fundamental Physics of the Universe, via Beirut 2, I-34151 Trieste, Italy \and
   Dipartimento di Matematica e Fisica, Universit\`a Roma Tre, Via della Vasca Navale 84, 00146 Roma, Italy \and
   INFN, Sezione Roma1, Dipartimento di Fisica, Universit\`a di Roma La Sapienza, Piazzale Aldo Moro 2, I-00185 Roma, Italy  \and
   Dipartimento di Fisica e Astronomia ‘Augusto Righi’, Universit\`a degli Studi
di Bologna, via P. Gobetti, 93/2, 40129 Bologna, Italy\and
    INAF-Osservatorio di Astrofisica e Scienza dello Spazio di Bologna, via
Piero Gobetti, 93/3, I-40129 Bologna, Italy \and
    Institut d’Astrophysique de Paris, Sorbonne Universit\'e, CNRS, UMR 7095, 98 bis bd Arago, 75014 Paris, France \and
    Dipartimento di Fisica, Sezione di Astronomia, Universit\`a di Trieste, via Tiepolo 11, I-34143 Trieste, Italy \and
    Centre for Extragalactic Astronomy, Department of Physics, Durham University, South Road, Durham DH1 3LE, UK \and
    Center for Astrophysics — Harvard \& Smithsonian, Cambridge, MA 02138, USA \and
    INFN - Sezione di Roma “Tor Vergata”, Via della Ricerca Scientifica 1, 00133 Roma, Italy \and
    Space Science Data Center, SSDC, ASI, Via del Politecnico snc, 00133 Roma, Italy \and
    Centro de Astrobiología (CAB), CSIC-INTA, Camino Bajo del Castillo s/n, ESAC campus, 28692 Villanueva de la Cañada, Spain \and
    NASA Goddard Space Flight Center, Greenbelt, MD 20771, USA \and
    Scuola Normale Superiore, Piazza dei Cavalieri 7, I-56126 Pisa, Italy \and
    INAF--Osservatorio Astronomico di Padova, 
Vicolo dell'Osservatorio 5, I-35122, Padova, Italy \and
    European Space Agency, ESTEC, Keplerlaan 1, 2201 AZ Noordwijk, The Netherlands \and
    ASI - Agenzia Spaziale Italiana, Via del Politecnico snc, I-00133 Roma, Italy \and
    Dipartimento di Fisica, Universit\`a di Roma “Tor Vergata”, Via della Ricerca Scientifica 1, 00133 Roma, Italy \and
    Department of Astronomy, University of Maryland, College Park, MD 20742, USA \and
    INAF - Istituto di Astrofisica Spaziale e Fisica Cosmica Milano, Via A. Corti
12, 20133 Milano, Italy \and
    INFN–National Institute for Nuclear Physics, via Valerio 2, I-34127 Trieste, Italy \and 
    DiSAT, Universit\`a degli Studi dell’Insubria, Via Valleggio 11, I-22100 Como, Italy \and
    INFN, Sezione di Milano-Bicocca, Piazza della Scienza 3, I-20126 Milano, Italy \and
    INAF, Osservatorio Astronomico di Brera, Via E. Bianchi 46, I-23807 Merate, Italy \and
    Cavendish Laboratory, University of Cambridge, 19 J. J. Thomson Ave., Cambridge CB3 0HE, UK \and
    Kavli Institute for Cosmology, University of Cambridge, Madingley Road, Cambridge CB3 0HA, UK \and
    Department of Physics \& Astronomy, University College London, Gower Street, London WC1E 6BT, UK \and
    Dipartimento di Fisica, Universit\`a di Roma La Sapienza, Piazzale Aldo Moro 2, I-00185 Roma, Italy \and
    Sapienza School for Advanced Studies, Viale Regina Elena 291, I-00161 Roma, Italy 
}
\abstract{
The existence of luminous quasars (QSO) at the Epoch of Reionization (EoR; i.e. $z>6$) powered by well grown supermassive black holes (SMBH) with  masses $\gtrsim10^9\rm~\msun$ challenges models of early SMBH formation and growth. To shed light on the nature of these sources we started a multiwavelength program based on a sample of 18 HYPerluminous quasars at the Epoch of ReionizatION (\hyp). These are the luminous QSOs whose SMBHs must have had the most rapid mass growth during the Universe first Gyr and, hence, acquired the largest mass at their respective epochs. In this paper we present the \hyp\ sample and report on the first of the three-years planned observations of the 2.4~Ms \xmm\ Multi-Year Heritage program on which \hyp\ is based. The goal of this program is to accurately characterise the X-ray nuclear properties of QSOs at the EoR. Through a joint X-ray spectral analysis of 10 sources, covering the rest-frame $\sim2-50$~keV energy range, we report a steep average photon index ($\Gamma\approx2.4\pm0.1$). No absorption is required at levels of $10^{21}-10^{22}~\nhi$. The measured average $\Gamma$  is inconsistent at $\geq4\sigma$ level with the canonical value ($\Gamma=1.8-2$) measured in QSO at $z<6$. Such a steep spectral slope is also significantly steeper than that reported in lower-z analog QSOs with similar luminosity or accretion rate, thus suggesting a genuine redshift evolution. 
Alternatively, we can interpret this result as the presence of a very low,  almost unreported at lower-z, energy cutoff $E_{cut}\approx20$~keV on a standard $\Gamma=1.9$ power-law. 
We also report on mild indications that, on average,  \hyp\ QSOs show higher levels of coronal soft X-rays at 2~keV compared to the accretion disc UV at $2500\AA$ than expected by lower-z AGN in the high-luminosity regime. We speculate that either a redshift-dependent coupling between the X-ray corona and accretion disk or intrinsically different coronal properties may account for the steepness of the X-ray spectral slope, especially in the presence of powerful winds. The reported steep slopes, if confirmed also in lower-luminosity regimes, may have a important impact on the design of next-generation X-ray facilities and future surveys aimed at investigating the early Universe. 

}


\maketitle
\section{Introduction}
Almost 300 spectroscopically confirmed quasars (QSOs) have been reported to date at $z\approx6-7.6$ \citep[][and references therein]{fan2022} during the Epoch of Reionization (EoR). 
They are powered by Supermassive Black-Holes (SMBHs)
with masses ($\mbh$) from $\sim10^{8}$~$\msun$ up to
$\sim10^{10}$~$\msun$ shining with high bolometric luminosities ($\lbol$) in the range $10^{46}-10^{48}~\lumcgs$ ($\sim10^{13}-10^{15}$~$\lsun$) close to the Eddington luminosity limit ($\lumedd$), i.e. with Eddington ratio $\ledd=\lbol/\lumedd\gtrsim 0.2$ \citep[e.g.][]{willott2010,mazzucchelli2017,shen2019}.

The mere existence of $\mbh$ as large as $\sim10^{9}$~$\msun$, at EoR, poses serious challenges to theoretical models aimed at explaining how these systems formed in less than 1~Gyr \citep{volonteri2010,johnson2016}.  
If the high-z SMBH progenitors form at $z\approx20-30$ \citep[e.g.][]{valiante2016}, it would be necessary to have seed BHs of at least  $\sim 1000-10000$~$\msun$ continuously growing at the Eddington rate for $\sim0.5-0.8~\rm Gyr$  (assuming a standard radiative efficiency, $\epsilon=0.1$) to reach the typical $\mbh$ reported in $z>6-7$ quasars \citep[see e.g.][]{wu2015,banados2018,yang2020,wang2021a}. This is challenging as it requires 
the uninterrupted availability of $\sim 10^9$~$\msun$ gas supply
throughout the $\sim$billion years of growth \citep[][]{johnson2007,milosavljevic2009}.
A continuous feeding at the observed $\ledd<1$
would instead imply an initial seed mass $>10^4\rm~\msun$ for the large majority of currently discovered $z>6$ QSO.

\begin{figure*}[ht]
   \begin{center}
    \includegraphics[width=0.46\textwidth, angle=0]{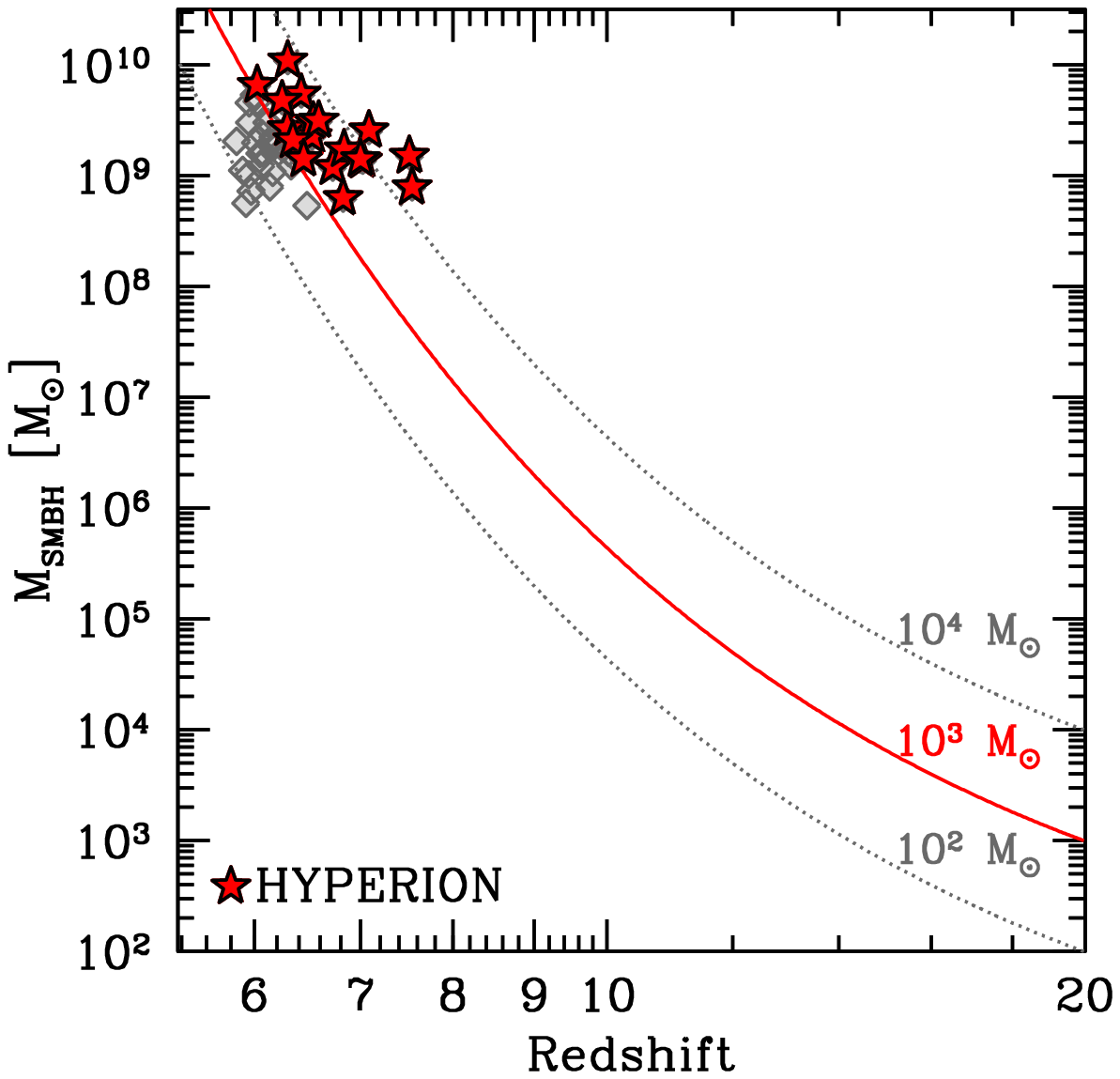}
    \includegraphics[width=0.46\textwidth, angle=0]{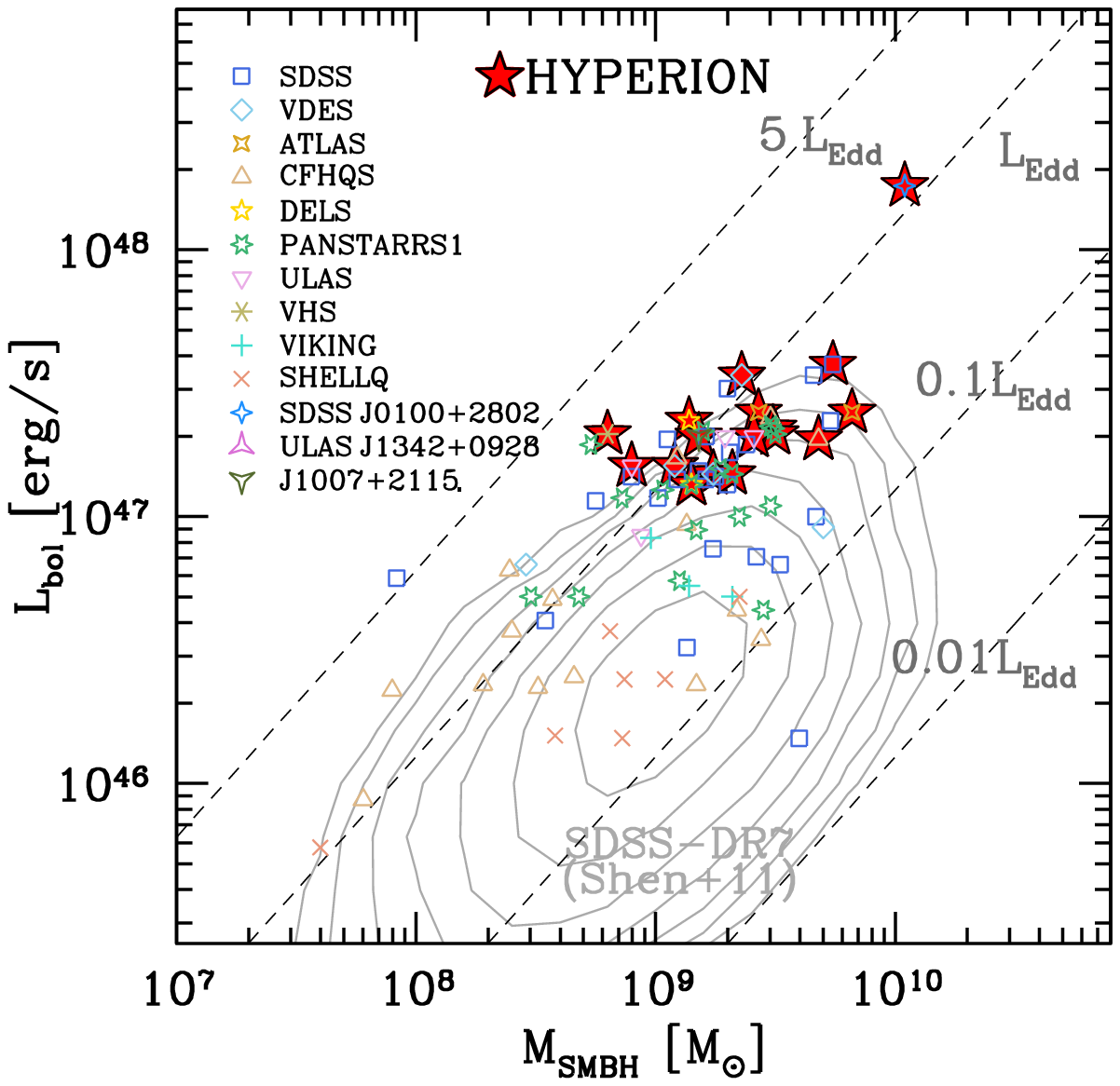}
    \end{center} 
    \caption{Selection and properties of the \hyp\ sample.  Left: SMBH mass as a function of redshift. All the reported points (diamonds and stars) are the sub-sample of 46 hyperluminous ($\lbol>10^{47}~\lumcgs$) quasars with known SMBH masses reported at the end of 2020. The final selected 18 sources are reported as red stars. The curves represent the exponential growth of seed BHs of different masses (labelled in figure) formed at $z=20$, assuming continuous accretion ($f_{duty}=1$) at the Eddington rate ($\ledd=1$; see Section~\ref{sample_definition}). The red curve, corresponding to a growing seed of 1000$~\msun$, has been used to select the \hyp\ sample.
    Right: Distribution of the \hyp\ sample in the $\mbh~vs~\lbol$ plane (red stars) along with the distribution of the 83 $z > 6$ quasars with available $\mbh$. All $\mbh$ are based on single-epoch MgII virial estimator  and $\lbol$ is from bolometric correction from literature as of 2020. $\mbh$ and $\lbol$ have been consistently recomputed for all sources assuming the same $\Lambda$CDM cosmology and adopting the mass calibration from \citet{vestergaard2009} and a bolometric  correction of 5.15 to the $3000$\AA\ luminosity from \citet{richards2006}. Dashed lines report the location of sources emitting a fixed fraction of $L_{\rm Edd}$. The contours report the location of the lower redshift ($z=0.7-1.9$) SDSS-DR7 quasars from \citet{shen2011} with MgII-derived masses.}
    \label{hyperion_sample}
\end{figure*}

Theoretical studies \citep[see][for recent reviews]{Inayoshi2020,lusso2023}, indeed, 
suggest that $z>6$ SMBHs must have formed from very large initial masses (i.e.  the so-called heavy seeds, $10^4-10^6$~$\msun$), 
growing at Eddington-limited gas accretion rates \citep[e.g.][]{volonteri2010,valiante2016}. Alternatively, they may have grown efficiently from lower-mass BHs ($\sim 100$~$\msun$; light seeds), remnants of Population III stars, through a series of short and intermittent super-Eddington accretion phases \citep[e.g.][]{volonteri2015,pezzulli2016}. However, the viability of these two channels of SMBH formation is still unclear \citep{johnson2016}. 
BHs  may  also  grow  through  coalescence  with  other  BHs  during  galaxy mergers, in the framework of the hierarchical structure formation scenario  \citep{volonteri2003,tanaka2009}. The merger time-scale of a binary BH is highly uncertain, but is likely to be long (up to few tens of Gyr) especially at high redshift  \citep[e.g.][]{biava2019,souzalima2020}.

The fundamental challenges posed by these sources have prompted a massive effort
in following-up the hyperluminous ($\lbol> 10^{47}~\lumcgs$) quasars at
near-infrared (UV/optical rest-frame) and far-infrared/sub-mm wavelengths with the largest and most sensitive observatories and with the best facilities available to
date \citep[e.g.][]{willott2010,wang2013,venemans2016,mazzucchelli2017,venemans2017,reed2017,feruglio2018,shen2019,onoue2019,fan2019,schindler2020,yang2021,farina2022,walter2022,bischetti2022}.
In the X-rays, despite similar dedicated observational efforts \citep[e.g.][]{brandt2002,farrah2004,moretti2014,page2014,gallerani2017,ai2017,nanni2018,banados2018b,pons2019,salvestrini2019,connor2020,wang2021,vito2021,vito2022,wang2022,wolf2023}, our
knowledge of the properties of $z>6$ quasars has been 
limited by the sensitivity and efficiency of current X-ray
observatories and the lack of all-sky X-ray suveys.
Despite this, few mostly marginal indications of different nuclear/host properties compared to lower-z QSO analogs have been reported. These indications involve: 1) hints of X-ray photon index steepening \citep[][]{vito2019,wang2021}; 2) faster/more frequent nuclear winds \citep[][]{meyer2019,schindler2020,yang2021,bischetti2022}; 3) $\mbh$ overgrown compared to host-galaxy dynamical mass with respect to the local scaling relation \citep[e.g.][]{pensabene2020,neeleman2021,farina2022,tripodi2023}. 
Given the challenging nature of massive $z>6$ QSOs, it is tempting to ascribe all those properties to their  
peculiar SMBH mass assembly history. 
However, their confirmation on firmer statistical grounds and their interpretation needs to be carefully evaluated. Indeed, an observational selection solely based on interesting, peculiar, bright sources, the availability of a restricted set of good quality data or the lack of a uniform physically-motivated sample selection may lead to a biased interpretation of these results.

The importance of an X-ray characterization of QSOs at EoR lies on the fact that the X-ray emission carries a nearly instantaneous information on the innermost active galactic nuclei (AGN) accreting regions. Indeed, a fraction of the thermal UV emission radiated by the accretion disc is reprocessed (i.e. Compton up-scattered) in the X-rays \citep[e.g.][]{haardt1993} by a compact inner \citep[i.e. 10-20 gravitational radii; e.g.][]{demarco2013,mcleod2015,chartas2016,kara2016} optically thin region, the hot corona. Such a radiation has a power-law spectral shape and a typical photon index  $\Gamma=1.8-2$ slope. The latter is constant up to $z\sim5$ \citep{piconcelli2005,vignali2005,shemmer2008,just2007,zappacosta2018}, falling off at high energies with an exponential cutoff at $>100$ keV \citep[e.g.][]{dadina2008,vasudevan2013,malizia2014,ricci2018}  depending on the physical properties of the corona \citep{fabian2015,fabian2017}. 
 The photon index has been claimed as a possible indicator of the AGN accretion rate as parameterized by the Eddington ratio $\ledd$, i.e. the mass-normalized bolometric luminosity (e.g. \citealt{shemmer2008,brightman2013,trakhtenbrot2017,liu2021}; but see \citealt{laurenti2022} and \citealt{kamraj2022}). A tight anti-correlation has long been reported between the accretion disc monochromatic UV luminosity at 2500~$\rm \AA$ ($\luvmon$) and the optical-to-X-ray spectral index ($\aox$), parametrizing the relative contributions of the accretion disc UV ($\luvmon$) and corona X-ray (2~keV; $\lxmon$) radiative outputs \citep[e.g.][]{avni1982,vignali2003,steffen2006,lusso2016,martocchia2017,timlin2020}. 
Physical properties and the relative geometrical configuration of the accretion disc/corona system therefore play a crucial role in shaping the $\aox$ and $\Gamma$ relations \citep[e.g.][]{kubota2018}. The validation of the $\aox$-$\luvmon$ relation at very high redshifts may allow us to extend and improve cosmology studies \citep{risaliti2019} up to those early epochs. 

Apart from the marginal indications of steeper $\Gamma$ in stacked/joint spectral fitting analysis of $z>6$ QSOs, past X-ray studies 
have not found other signs of convincing evolutionary properties. However,  they suffered from (i) limited constraining power due to low X-ray counts statistics ($<10-15$~net-counts),  preventing proper spectral analysis on source-by-source basis, and (ii) a small number of sources with reliable spectral data quality \citep[][]{nanni2017,ai2017,gallerani2017,nanni2018,vito2019,pons2020,wang2021,medvedev2021,wolf2023,connor2020}.

\begin{table*}[h!]
  \begin{center}
  \caption{The \hyp\ QSO sample, ordered by decreasing redshift, and its general properties}  
    \begin{tabular}{lrrrrrrrrc}
      \hline
      \hline
      \multicolumn{1}{c}{Name}      &   \multicolumn{1}{c}{RA}     &   \multicolumn{1}{c}{DEC}   &\multicolumn{1}{c}{$z^a$} & $\log \lbol$$^b$ & $\log \mbh$$^c$ &$\ledd$ & \multicolumn{1}{c}{$\rm M^{seed}_{BH}$} & \multicolumn{1}{c}{$\rm L_{2500}$$^d$} & Ref.$^e$\\
                &          &         &      &$\lumcgs$  & $\msun$  &      &  \multicolumn{1}{c}{$\msun$}     &  \multicolumn{1}{c}{$\lumcgs$} &  \\
      \hline	
ULAS~J1342+0928 & 13:42:08.10  &  +09:28:38.6  &7.541 &   47.19  &   8.90   & 1.55 &  19120 & $ 46.58\pm 0.02 $   &    1\\ 
    J1007+2115  & 10:07:58.26  &  +21:15:29.2  &7.494 &   47.30  &   9.18   & 1.05 &  32460 & $ 46.66\pm 0.03 $   &    2\\ 
ULAS~J1120+0641 & 11:20:01.48  &  +06:41:24.3  &7.087 &   47.30  &   9.41   & 0.62 &  18230 & $ 46.71\pm 0.07 $   &    3\\ 
DELS~J0038-1527 & 00:38:36.10  &  -15:27:23.6  &7.021 &   47.36  &   9.14   & 1.32 &   7983 & $ 46.79\pm 0.04 $   &    4\\ 
DES~J0252-0503  & 02:52:16.64  &  -05:03:31.8  &6.99  &   47.12  &   9.15   & 0.74 &   7679 & $ 46.55\pm 0.04 $   &    5, 6\\
VDES~J0020-3653 & 00:20:31.47  &  -36:53:41.8  &6.834 &   47.16  &   9.24   & 0.66 &   5753 & $ 46.64\pm 0.05 $   &    7\\ 
 VHS~J0411-0907 & 04:11:28.62  &  -09:07:49.7  &6.824 &   47.31  &   8.80   & 2.57 &   2019 & $ 46.71\pm 0.03 $   & 	  8\\ 
VDES~J0244-5008 & 02:44:01.02  &  -50:08:53.7  &6.724 &   47.19  &   9.08   & 1.02 &   2814 & $ 46.55\pm 0.03 $   &    7\\ 
PSO~J231.6-20.8 & 15:26:37.84  &  -20:50:00.7  &6.587 &   47.31  &   9.50   & 0.51 &   4708 & $ 46.66\pm 0.06 $   &    9\\ 
PSO~J036.5+03.0 & 02:26:01.88  &  +03:02:59.4  &6.533 &   47.33  &   9.49   & 0.55 &   3776 & $ 46.78\pm 0.03 $   &    9\\ 
VDES~J0224-4711 & 02:24:26.54  &  -47:11:29.4  &6.526 &   47.53  &   9.36   & 1.18 &   2730 & $ 46.83\pm 0.04 $   &    7\\ 
PSO~J011+09     & 00:45:33.57  &  +09:01:56.9  &6.444 &   47.12  &   9.15   & 0.74 &   1279 & $ 46.37\pm 0.02 $   &    9\\ 
SDSS~J1148+5251 & 11:48:16.64  &  +52:51:50.2  &6.422 &   47.57  &   9.74   & 0.54 &   4627 & $ 46.90\pm 0.02 $   &    10\\ 
PSO~J083.8+11.8 & 05:35:20.90  &  +11:50:53.6  &6.346 &   47.16  &   9.32   & 0.55 &   1324 & $ 46.69\pm 0.03 $   &   11\\ 
SDSS~J0100+2802 & 01:00:13.02  &  +28:02:25.8  &6.300 &   48.24  &  10.04   & 1.26 &   5799 & $ 47.56\pm 0.07 $   &   12\\ 
ATLAS~J025-33  &  01:42:43.70  & -33:27:45.7   &6.294 &   47.39  &   9.57   & 0.72 &   1392 & $ 46.93\pm 0.01 $   &  13 \\
CFHQS~J0050+3445& 00:50:06.67  &  +34:45:22.6  &6.246 &   47.29  &   9.68   & 0.32 &   2072 & $ 46.67\pm 0.03 $   &    10\\ 
ATLAS~J029-36   & 01:59:57.97  &  -36:33:56.6  &6.027 &   47.39  &   9.82   & 0.30 &   1220 & $ 46.60\pm 0.03 $   &   13\\   
\hline
    \end{tabular}
  \label{sample}
  \end{center}
{\footnotesize $^a$: measured from the MgII emission line; $^b$: estimated from luminosity $3000\rm\AA$ ($\rm L_{3000~\rm\AA}$, see reference column) from \citet{richards2006}; $^c$: measured from single epoch virial mass estimator employing the FWHM of the MgII line and $\rm L_{3000~\rm\AA}$ from \citet{vestergaard2009}; $^d$: estimated through interpolation of adjacent photometric points (Saccheo et al. in prep.); $^e$: References for redshift and parameters to estimate $\lbol$ and $\mbh$: 
1. \citet{banados2018}; 
2. \citet{yang2020}; 
3. \citet{mortlock2011}; 
4. \citet{wang2018}; 
5. \citet{wang2020}; 
6. \citet{yang2021}; 
7. \citet{reed2019}, 
8. \citet{pons2019}; 
9. \citet{mazzucchelli2017}; 
10. \citet{shen2019};
11. \citet{andika2020};
12. \citet{wu2015};
13. \citet{chehade2018}
\vspace{-0.5cm}
}
\end{table*}

In this paper we present 1) a new sample of $z>6$ QSOs selected with a physically motivated criterion  to include the titans among $z>6$ QSOs: i.e. those powered by SMBHs which appear to have undergone the fastest BH growth compared to other co-eval sources; 
2) a \xmm\ Multi-Year Heritage X-ray program on this sample designed to begin the first systematic X-ray spectroscopic exploration of QSOs at EoR; 3) the results of the first year of the \xmm\ program. 
In Section~\ref{sample_definition} we present our QSO sample and the \xmm\ Multi-Year Heritage X-ray program. The reduction of the X-ray data from the 1st year of the Heritage program and X-ray photometry is described in Section~\ref{datareduction}. The X-ray spectral analysis is reported in Section~\ref{specanalysis}. The results, their discussion  and relative conclusions are given in Section~\ref{results}, Section~\ref{discussion} and Section~\ref{conclusions}, respectively. 

Throughout the paper we adopt a $\Lambda C\rm DM$ cosmology with $ H_0=\, 70 \rm~km\,s^{-1}\,Mpc^{-1}$, $\Omega_M=0.27$ and $\Omega_\Lambda=0.73$. Errors are reported at $1\sigma$ level with upper limits quoted at 90\% confidence level.

\section{The \hyp\ Sample and the \xmm\ Heritage Program}\label{sample_definition}
The HYPerluminous quasars at the Epoch of ReionizatION (\hyp) sample is defined by the selection of all the $z>6$ hyperluminous QSOs ($\lbol\geq10^{47}~\rm \lumcgs$) known up to 2020 which required an initial seed BH mass of $\mseed>1000~\rm M_\odot$ accreting via continuous exponential growth at the Eddington rate to form the measured SMBH mass. The selection was performed on the 46 unlensed radio-quiet hyperluminous $z>6$ QSOs known with published SMBH masses at the end of 2020 \citep[i.e.][]{willott2010,derosa2011,mazzucchelli2017,wu2015,banados2018,shao2017,reed2019,wang2018,pons2019,chehade2018,shen2019,yang2020,wang2020,eilers2020,andika2020,onoue2019,matsuoka2019}. 
The selection criterion of the \hyp\ QSOs is reported in Fig.~\ref{hyperion_sample} as the red curve. The curves represent the time-dependent exponential mass growth modelled as $M_{\rm BH}=\mseed\times \exp{(t/t_s)}$ with e-folding time $t_s=0.45 \, \epsilon\,(1-\epsilon)^{-1}\,\ledd^{-1}\,f_{duty}^{-1}\,\rm Gyr$
of seed BHs of different masses (labelled in figure) formed at $z=20$ \citep[][]{valiante2016}, assuming continuous accretion at the Eddington rate, i.e. $\ledd=1$, radiative efficiency $\epsilon=0.1$ and active phase duty cycle $f_{duty}=1$.
Hence this sample includes the "titans" among QSOs, i.e. those powered by the SMBH which had the largest mass assembly over the Universe first Gyr. 

Notice that this selection criterion is a convenient way to statistically select the sample of QSOs which are powered by SMBH which experienced the most rapid growth during their formation hystory.
  This selection is physically motivated as it allows to identify these sources through a reference curve, starting at a specific $\mseed$, for the continuous Eddington-limited mass growth. Under this assumption the $\mseed$ reported in Table~\ref{sample} and required by each SMBH to grow its mass has to be considered solely as a proxy for the mass growth rate experienced by each SMBH and not necessarily a physically meaningful quantity.

All \hyp\ QSOs have been selected in optical-to-mid infrared and benefit from extensive high-quality multi-band photometric/spectroscopic coverage from rest-frame UV (i.e. observed NIR band) to sub-mm/mm band. 
By definition, NIR spectroscopic data from VLT, Magellan, Gemini, Keck spectrographs are available for all the \hyp\ QSOs. From these data MgII-based single epoch virial masses and bolometric luminosities  from $3000$~$\AA$ bolometric correction have been derived \citep[e.g.][]{wu2015,mazzucchelli2017,banados2018,reed2019,shen2019}. Similarly, photometric data, especially for NIR and sub-mm/mm, are available to different quality levels \citep[e.g.][Saccheo et al. in prep.]{tripodi2023,feruglio2023}. 

We have obtained $\mbh$ employing the MgII virial mass estimator by \citet{vestergaard2009} which employs the full-width at half maximum (FWHM) of the MgII line and the $3000\rm~\AA$ continuum luminosity. We also computed $\lbol$ via the  $3000\rm~\AA$ bolometric correction from \citet{richards2006}. Notice that this choice of virial mass estimator makes our selection conservative and therefore robust as among the MgII virial mass estimators, the one from \citet{vestergaard2009} tends to give the lowest SMBH mass estimates \citep[see e.g.][]{farina2022} and hence the lowest $\mseed$. Furthermore, the average E(B-V) estimated trough a Spectral Energy Distribution (SED) analysis for the \hyp\ QSOs is $<0.01$ (Saccheo et al. in prep.) and hence the mass estimates are not affected by spectral reddening. 
Fig.~\ref{hyperion_sample} right shows the distribution of the \hyp\ QSOs in the $\mbh~vs.~\lbol$ plane along with the distribution of the 83 $z>6$ QSOs with estimated masses known by the end of 2020. 
The \hyp\ QSOs, distributed in the redshift range $z\approx6-7.5$ (mean $z\sim6.7$),  have an average $\log(L_{\rm bol}/\rm
erg\,s^{-1})\approx47.3$ and span a mass range $\approx 10^{9}-10^{10}~\msun$ leading to $\ledd=0.3-2.6$. 
Notice that the virial mass estimates uncertainties are dominated by systematics reaching 0.3-0.5~dex \citep[e.g.][]{shen2012}. To have a sense of the variation of the estimated masses, if we employs for our selection the \citet{shen2011} MgII-based mass estimator which typically gives high $\mbh$ estimates \citep{farina2022}, we obtain SMBH masses 0.2~dex higher implying $\ledd$ smaller by $\sim40$\% and 
 $\rm M^{seed}_{BH}$ larger by a factor 1.6.

Table~\ref{sample} lists the 18 selected QSOs in the \hyp\ sample along with their celestial coordinates, MgII-based redshifts, $\lbol$, $\mbh$, Eddington ratio ($\ledd$), required $\mseed$ and $\luvmon$ obtained trough interpolation of photometric points (Saccheo et al. in prep.). All the quantities have been re-evaluated by uniformly adopting a $\Lambda$CDM cosmology with $\Omega_M=0.27$ and $\Omega_\Lambda=0.73$. 
Hereafter, we refer to the single QSOs shortening their name, reported in Table~\ref{sample}, as J plus the digits of their RA.

For two \hyp\ QSOs (J0224 and J0100) good quality archive X-ray data from \xmm\ are already available and their spectral analysis has been presented in \citet{pons2019} and \citet{ai2017}.
For the remaining \hyp\ QSOs we have an on-going 2.4~Ms \xmm\ Multi-Year Heritage Program (PI L. Zappacosta; Proposal ID 088499), approved in Dec~2020 with a three years time span, to collect unprecedented high quality X-ray data for such a large sample of QSOs at EoR. Specifically, the \hyp\ \xmm\ Multi-Year Heritage Program (hereafter \hyph) is collecting for the first time X-ray data for seven sources and is improving the data quality for already observed nine sources but for which only limited X-ray data quality \citep[either leading to non-detections or to mainly 10-15 net-counts detections, e.g.][]{vito2019,pons2019,connor2020} is available. 
The aim of \hyph\ is to achieve for all QSOs in the sample the high-quality data standard that up to now has been obtained in unlensed QSOs by J0224 and J0100 (i.e. at least 100 net-counts from pn+MOS1+MOS2 data in the 0.5-10~keV band).
This would ensure a $\sim$10\% accuracy level (1$\sigma$) characterization of the X-ray spectral properties photon index of the power-law  and unabsorbed 2-10~keV luminosity ($\lumh$) on these sources.

\begin{figure*}[t!]
   \begin{center}
    \includegraphics[width=0.95\textwidth, angle=0]{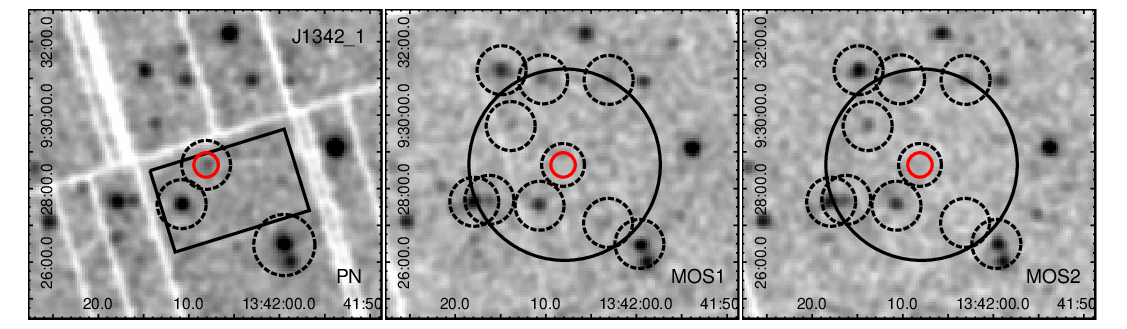}
   \end{center} 
\caption{EPIC 0.5-2~keV pn, MOS1 and MOS2 camera images for the first J1342 observation reported in Table~\ref{obsjournal} of the \hyph\ program presented in this work. All the images are smoothed by a Gaussian kernel of 3~pixel radius for better visualization. Source and background counts/spectral extraction regions are reported in red and black, respectively. Dashed circular regions indicates areas excluded from the background extraction. Images for the second J1342 exposure ($J1342\_2$) and other \hyph\ QSOs presented in this work are reported in Appendix~\ref{hyperion_subsample_others}.}
       \label{hyperion_subsample}
\end{figure*}

\setlength{\tabcolsep}{11pt}
\begin{table*}[t!]
    \caption{Journal of the observations of the \hyp\ targets from the \hyph\ (upper part of the table) and archive (lower part of the table).}
    \begin{center}
    \begin{tabular}{lccrrrrrr}
    \hline
    Source  &  OBSID$^{a}$  & Start date   & \multicolumn{3}{c}{Nominal Exposure (ks)}           &  \multicolumn{3}{c}{Cleaned exposure (ks)} \\
            &                 &              & pn&MOS1&MOS2  &  pn&MOS1&MOS2  \\
    \hline
    \multicolumn{9}{c}{\hyp\ \xmm\ Heritage program} \\
$\rm J1342\_1^{b}$  &  0884990101  & 2021-07-05 18:18:38 &106.5&99.7&97.7  &  60.3&78.5&78.4 \\
$\rm J1342\_2^{b}$  &  0884993801  & 2021-12-24 12:49:15 &101.5&98.4&102.6 &  46.4&68.2&73.8 \\
J1120  &  0884990401  & 2021-06-27 18:30:48 & 71.6&73.4&73.4    &  36.6&51.5&52.6 \\
J0020  &  0884991101  & 2022-01-01 05:50:36 & 85.8&87.6&87.6  &  37.4&62.0&63.7 \\
J0244  &  0884991501  & 2021-08-04 17:03:07 & 87.2&89.0&89.0  &  67.4&81.3&78.5 \\
J231.6  &  0884991701   & 2021-07-29 17:30:46 & 109.5&108.0&106.6  &  66.9&89.0&89.9 \\
J036.5  &  0884992001   & 2021-07-19 18:11:14 & 84.8&74.6&70.9  &  47.1&68.0&66.3 \\
J011  &  0884992101   & 2021-07-15 18:14:32 & 81.3&76.9&72.1  &  46.2&56.0&58.4 \\
J083.8   &  0884992401  & 2022-03-14 00:46:23 & 84.8&86.6&86.6  &  51.9&74.0&73.6 \\
J0050  &  0884992601  & 2021-06-26 18:27:30 & 42.8&44.6&42.8  &  26.2&32.9&34.5 \\
J029   &  0884992901  & 2022-01-03 17:05:23 & 84.0&84.5&85.1  &  55.2&69.2&68.6 \\
\hline
    \multicolumn{9}{c}{\hyp\ archival observations}\\
J0224 &  0824400301   & 2018-05-25 11:35:29 & 32.7&34.5&34.5  &  14.9&23.8&27.4 \\
J0100 &  0790180701   & 2016-06-29 17:53:42 & 62.4&64.1&64.0  &  41.1&60.1&55.9 \\
\hline
    \end{tabular}
    \label{obsjournal}
    \end{center}
    {\footnotesize $^a$: observation ID for each \xmm\ dataset considered; $^b$: suffix $\rm \_1$ and $\rm \_2$ refer to the first and second exposure for source J1342.
    }
\end{table*}

\section{Data reduction and photometry}\label{datareduction}
In this work we report data from the first year of observations of the \hyph\ program. 
In addition, we also perform a re-analysis of the two archival \hyp\ sources J0224 and J0100, for consistency.
In total we are presenting $\sim0.94$~Ms of new data on ten sources which increases to $\sim1.04$~Ms accounting for the observations of the two archival sources. 
Table~\ref{obsjournal} presents the details on the considered observations. 
\hyph\ observations have been already completed, with two exposures, for only one out of all the considered targets, i.e. J1342.
Observations for the remaining targets in this sub-sample will be completed over the following two years of the \hyph\ program with at least one further exposure. The exact schedule of the exposures is flexible and may vary depending on the flux state of each target measured on their first exposure.  

The \xmm\ data have been processed with the  SAS v19.1.0. Following the standard procedures outlined in the \xmm\ science threads we created through the {\em epicproc} package newly calibrated event files. We produced the high energy light curves for the EPIC pn and MOS detectors in the energy range 10-12~keV and $>10~\rm keV$, respectively. We visually inspected them for the presence of high background flares. Following the recommendations presented in the most updated calibration technical notes\footnote{XMM-SOC-CAL-TN-0018 which is available at https://xmmweb.esac.esa.int/docs/documents/CAL-TN-0018.pdf}, we identified the good time intervals by removing the part of the pn observations which were affected by rates higher than 0.4~cts/s ($\sim0.41$~cts/s for J011). As for the MOS1 and MOS2 exposures, we adopted thresholds in the range 0.12-0.17~cts/s and 0.18-0.22~cts/s, respectively. 
We determined that for the pn only one observation (J0244) had $\sim20$\% of the exposure affected by high background periods. All other sources had their observation impacted by $\sim40-50$\%. As for the MOS exposures, we calculated  a percentage of time affected by high backgrounds in the range $\sim8-27$\% and $\sim6-19$\% for MOS1 and MOS2, respectively. In Table~\ref{obsjournal} we report nominal and cleaned exposures for each observation.

\setlength{\tabcolsep}{8pt}
\begin{table*}[t!]
    \caption{Source net-counts from fixed-aperture photometry on the EPIC detectors in the soft, hard and full energy bands}
    \begin{center}
    \begin{tabular}{lccccccccc}
    \hline
    Source  & \multicolumn{3}{c}{$\rm Counts_{0.5-2~keV}$(cts)}  & \multicolumn{3}{c}{$\rm Counts_{2-10~keV}$(cts)} & \multicolumn{3}{c}{$\rm Counts_{0.5-10~keV}$(cts)}\\
            &                 pn&MOS1&MOS2  &  pn&MOS1&MOS2  &  pn&MOS1&MOS2  \\
    \hline
        \multicolumn{10}{c}{\hyp\ \xmm\ Heritage program} \\
$\rm J1342\_1^{a}$  &   $41.2_{-12.3}^{+13.0}$  & $<17.6$                &$<17.2$                  & $<18.6$           &$<7.6$               & $<13.3$           & $<60.7$    &$<15.1$                 &$<22.5$ \\
$\rm J1342\_2^{a}$  &   $29.7_{-11.1}^{+11.9}$  & $<19.1$                & $18.3_{-7.7}^{+8.3}$    & $<17.6$           & $<9.4$              & $<10.1$           & $<50.2$                   & $<19.1$                & $<29.7$  \\
J1120  &  $33.6_{-8.6}^{+9.2}$  &  $18.3_{-6.3}^{+7.1}$  &  $20.7_{-6.6}^{+7.1}$      & $<13.6$           & $<15.1$             & $<11.2$           & $29.4_{-12.2}^{+12.8}$      & $23.4_{-8.9}^{+9.5}$        &  $21.3_{-8.6}^{+9.3}$  \\
J0020  &   $14.1_{- 5.6}^{+ 6.2}$  & $<11.5$                & $<13.7$                 & $<12.0$           & $ <13.3$            & $<5.1$            & $<28.1$                   & $<19.9$                & $<12.2$  \\
J0244  &   $59.0_{-10.0}^{+10.7}$  & $11.9_{-5.2}^{+5.9}$   &$  26.7_{-6.3}^{+7.0}$   &$  <24.6        $  &$       <8.3     $&$         <17.1    $&$   68.9_{-13.5}^{+14.2}$   &$     <21.5          $&$   34.3_{-8.2}^{+8.9}$   \\
J231.6   &   $    <40.6        $   & $   <21.7        $    &$  <25.4           $    &$   <24.7      $  &$         <24.2  $&$           <10.7   $&$         <52.1      $   &$      <38.0          $&$      <26.2   $        \\
J036.5   &   $24.9_{-9.2}^{+9.7}$    & $   <16.9       $     &$  <17.4           $    &$   <17.1      $  &$         <21.4  $&$           <22.4   $&$         <41.9      $   &$      <31.4          $&$      <33.2   $        \\
J011   &   $    <26.1        $   & $   <13.6        $    &$  <9.6            $    &$   <18.1      $   &$        <16.6  $&$           <8.7    $&$         <33.3      $    &$     <23.8          $&$      <12.0   $         \\
J083.8   & $40.9_{- 10.1}^{+ 10.7}$ &  $<14.5$             & $<21.4$                & $<41.2$           & $19.3_{- 7.8}^{+ 8.6}$                &  $<26.0$                 & $65.1_{- 15.7}^{+ 16.4}$ & $23.6_{- 10.0}^{+ 10.6}$  & $26.4_{- 9.7}^{+ 10.4}$  \\
J0050  &   $16.9_{-7.0}^{+7.7}$    & $16.1_{-5.4}^{+6.0}$  &$   11.3_{-4.9}^{+5.7}$   &$  <14.1       $  &$        <16.7   $&$          <17.7    $&$        <32.8       $   &$     24.6_{-7.4}^{+8.1}$ &$    20.6_{-7.3}^{+8.0}$  \\
J029   &  $<32.5 $                 & $<18.9$               & $<14.1$                & $<37.2$            & $<7.1$              & $<12.4$                 & $39.9_{-14.6}^{+ 15.3}$  &  $<15.9$            & $<19.6$             \\
\hline
    \multicolumn{10}{c}{\hyp\ archival observations}\\
J0224  &   $45.5_{-8.3}^{+8.9}$    & $18.0_{-5.6}^{+6.1}$  &$   21.5_{-5.9}^{+6.6}$   &$  <22.1       $  &$        <12.5   $&$          <12.7   $&$    57.1_{-10.8}^{+11.5}$   &$    22.8_{-7.1}^{+7.9}$ &$    26.4_{-7.5}^{+8.3}$  \\
J0100  &  $157.9_{-14.1}^{+14.7}$  & $77.9_{-9.7}^{+10.3}$ &$    51.6_{-8.2}^{+8.8}$  &$   43.2_{-11.0}^{+11.8}$ &$    <19.8 $&$            <14.0 $&$     201.1_{-17.9}^{+18.7}$ &$   88.2_{-11.5}^{+12.1}$ &$   56.7_{-9.8}^{+10.5}$ \\
\hline
    \end{tabular}
    \label{counts}
    \end{center}
        {\footnotesize $^a$: suffix $\rm \_1$ and $\rm \_2$ refer to the first and second exposure for source J1342.
    }
\end{table*}

We first identified the point-like sources across the field of view. We created 0.5-2~keV energy band pn images and run on them the task {\em edetect\_chain} by setting a detection maximum likelihood (DETML\footnote{DETML$=-\ln P_{rnd}$, where $P_{rnd}$ is the probability of detection by chance.}) threshold $\rm DETML=6$. 
This blind search also produced the detection of all the QSO targets, except J011, J0020 and J231.6 (but see later for J0020). 
We verified the targets detection, accounting for the source position prior and performing forced aperture photometry on the QSO positions. We extracted the source counts on circular regions of radius 20~arcsec (corresponding to $\sim80$\% of the on-axis PSF encircled energy fraction at 1.5~keV) centered on the QSO optical position (see Table~\ref{hyperion_sample}), except for J0244 and J0020 which had a nearby source (28~arcsec and 17~arcsec distant), for which we adopted smaller apertures of 15~arcsec and 12~arcsec radius ($\sim65-70$\% of the PSF encircled energy fraction), respectively.
The background counts were extracted for the MOS cameras on circular apertures of radius in the range 2.5-3.4~arcmin centered on the QSO position. For the pn camera we adopted rectangular regions positioned around the target, rotated roughly with the same detector position angle and with long and short sides in the range 3.6-3.9~arcmin and 1.9-2.7~arcmin, respectively. 
The background counts extraction was performed excluding detector circular regions of 40~arcsec radius centered on (1) the contaminant point sources previously identified on the 0.5-2~keV pn image, (2) other sources reported on both MOS cameras and (3) the target QSOs. 
In case of bright sources we excluded larger regions of 50~arcsec radius. Fig.~\ref{hyperion_subsample} shows the adopted extraction regions on the 0.5-2~keV images of the three \xmm\ cameras for the first observation of J1342 ($\rm J1342\_1$; see Table~\ref{obsjournal}). Images and adopted extraction regions for the second observation of J1342 ($\rm J1342\_2$) and other sources are reported in Appendix~\ref{hyperion_subsample_others}. 

We computed the $\geq99$\% confidence level source detection by calculating the no-source binomial probability and estimated net-counts (with uncertainties) on the 0.5-2~keV (soft band), 2-10~keV (hard band) and 0.5-10~keV (full band) images of the three \xmm\ cameras \citep[see][]{weisskopf2007,vito2019}. 
We considered as detections the sources with a no-source binomial probability $\leq1$\% on either the pn detector or in both the MOS detectors in at least one band. Table~\ref{counts} reports the measured source counts with uncertainties. All the targets  resulted in detections except J011 and J231.6 in broad agreement (J0020 is detected in this case) with the results of source detection search performed across the field.
For the sources J083.8 and J029 this is the first X-ray detection reported.

For the undetected sources, we calculated pn upper limits on fluxes, luminosities and on $\aox$\footnote{We adopted as a reference the pn detector which is the most efficient detector at $<2$~keV energy where most detections occur.}.  They are reported in Table~\ref{analysis} along with the spectral measurements for the detected sources (see Section~\ref{specanalysis}). Specifically, we estimated the total counts in the soft and hard band by correcting the fixed aperture photometry of 20~arcsec radius reported in Table~\ref{counts} and accounting for $\sim80\%$ of the 1.5~keV total encircled energy fraction. 
We estimated the fluxes using the X-ray spectral fitting package XSPEC \citep{arnaud1996} assuming the spectral response files extracted at the source position and adopting a power-law model with both $\Gamma=2$ and $\Gamma=2.4$ (i.e. the average $\Gamma$ from a joint spectral analysis of the detected sources, see Section~\ref{jointspec} for details) absorbed by the Galactic column density. The latter is taken from the HI4PI survey \citep{HI4PI2016} as weighted average at the position of each source within a radius of 0.1~deg.  We then estimated the unabsorbed 2-10~keV and 2~keV luminosities with XSPEC by assuming the same absorbed power-law spectral model. 
\begin{figure*}[!]
   \begin{center}
    \includegraphics[width=0.72\textwidth, angle=0]{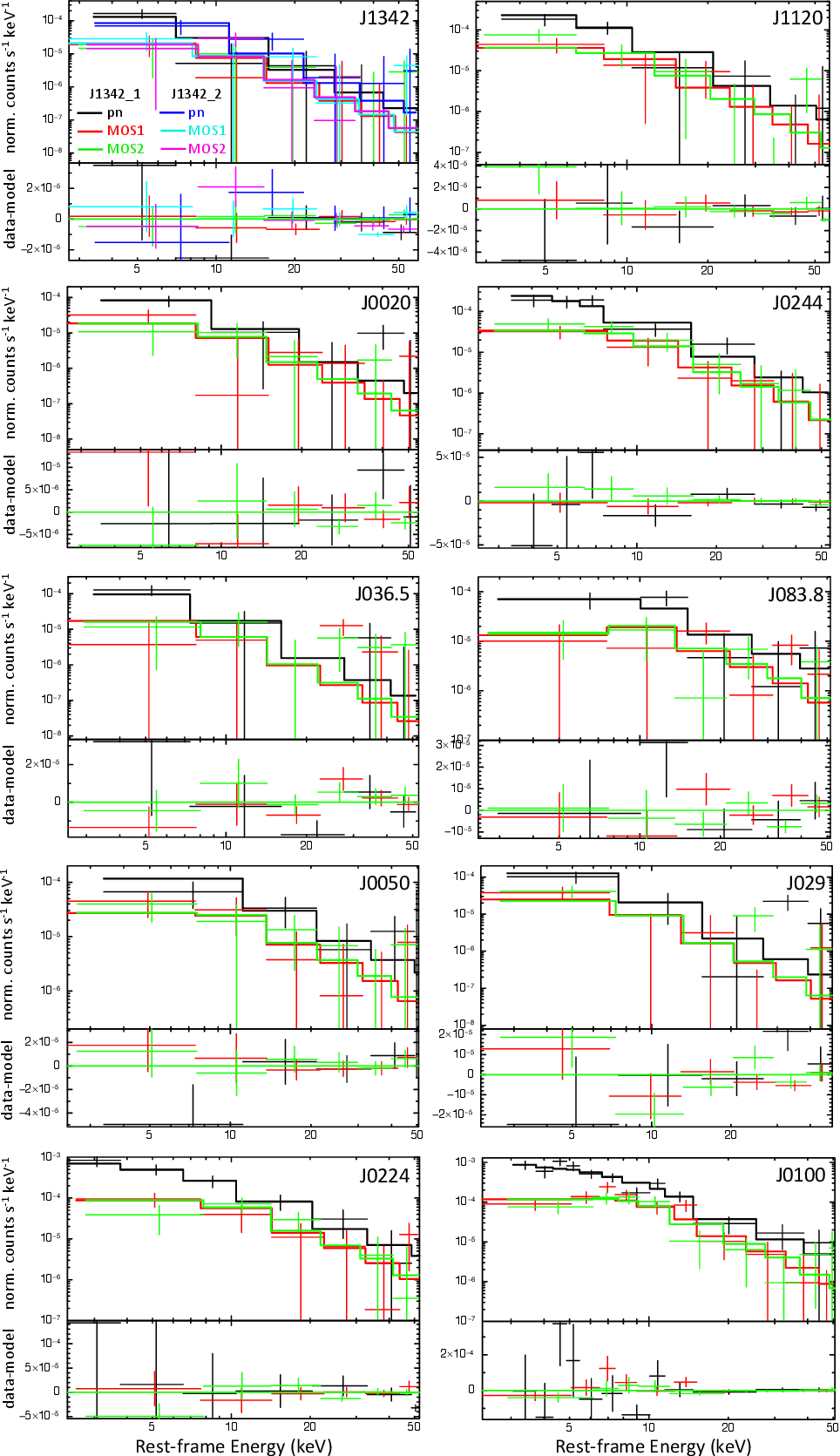}
   \end{center} 
\caption{\xmm\ pn (black),  MOS1 (red) and MOS2 (green)  0.3-7~keV spectra and best-fit models (stepped continuous thick lines) for the 10 detected \hyp\ QSOs presented in this paper. Spectra have been further rebinned for visual purposes and are reported at their rest-frame energy. Residuals are shown as data minus best-fit model in the bottom panels. For the source J1342 blue, cyan and magenta represent the second data set. }
       \label{spectra}
\end{figure*}
\section{Spectral analysis}\label{specanalysis}
In the following, we report the spectral analysis performed for all the detected sources in Table~\ref{counts}.
\subsection{Single source analysis}
The source and background spectral extractions were performed on the same regions adopted for the counts extractions. 
Given the sources low-counts and background dominated regime, before performing the spectral analysis we evaluated the best spectral binning scheme. We  simulated different input spectra and evaluated the accuracy of each binning scheme in recovering the input power-law parameters. We tried the following binning schemes:  minimum 1, 3, 5, 10 counts per bin and the optimal \citet[][KB hereafter]{kaastra2016} grouping. We verified that the KB binning\footnote{For this particular binning we used the FTOOLS (http://heasarc.gsfc.nasa.gov/ftools) command ftgrouppha with grouptype option "opt".}, which provides the optimal binning for data and model accounting for the source spectral shape, the variable spectral resolution and the average photon energy in each bin, is the best scheme to recover unbiased estimates of the parameters, and is also insensitive to the energy over which the spectral analysis is performed. We use the KB scheme for the following  spectral analysis. See Appendix~\ref{optimalbin} for a detailed description of the simulations. 

The spectral analysis was performed with XSPEC v12.11.1. We performed the modelings by using the Cash statistics with direct background subtraction \citep[W-stat in XSPEC;][]{cash1979,wachter1979}. 
\setlength{\tabcolsep}{5pt}
\begin{table*}[t!]
    \caption{Best-fit parameters from the X-ray spectral analysis.}
\centering
    \begin{tabular}{lcrcccccc}
    \hline
    \hline
Source &  W-stat/dof & Counts$^a$ &$\Gamma$           & $F_{0.5-2}$                 &   $F_{2-10}$             &     $L_{2-10}$        &     $L_{2~keV}$             &  $\aox$     \\
       &             &  pn/MOS$^b$       &             & $10^{-16}\,\fluxcgs$  & $10^{-16}\,\fluxcgs$   & $10^{44}\,\lumcgs$ & $10^{44}\,\lumcgs$ & \\
    \hline
J1342 & 392.7/319 &   75/  48 & $2.87_{-0.37}^{+0.43}$ & $ 10.54_{ -1.81}^{+  1.84}$ & $  3.65_{ -0.64}^{+  0.64}$ & $17.09_{-2.96}^{+3.05}$  & $9.86_{-1.71}^{+1.76}$ & $-1.48_{-0.03}^{+0.03}$   \\ 
J1120 & 177.0/157 &   42/  46 & $2.59_{-0.32}^{+0.35}$ & $ 21.59_{ -3.30}^{+  3.35}$ & $ 12.26_{ -1.89}^{+  1.88}$ & $26.71_{-3.96}^{+4.16}$  & $12.88_{-1.91}^{+2.00}$ & $-1.50_{-0.02}^{+0.03}$  \\ 
J0020 & 156.0/153 &   27/  20 & $2.75_{-0.53}^{+0.59}$ & $ 11.62_{ -2.46}^{+  2.47}$ & $  4.64_{ -1.00}^{+  0.98}$ & $12.64_{-2.60}^{+2.81}$  & $6.78_{-1.40}^{+1.51}$ & $-1.58_{-0.03}^{+0.04}$   \\ 
J0244 & 147.5/157 &   80/  62 & $2.39_{-0.22}^{+0.24}$ & $ 21.53_{ -2.32}^{+  2.37}$ & $ 15.23_{ -1.66}^{+  1.62}$ & $18.83_{-1.99}^{+2.08}$  & $7.87_{-0.83}^{+0.87}$ & $-1.52_{-0.02}^{+0.02}$   \\ 
J231.6 &   --      & --           & 2.0$^c$                   & $< 13.56^d$                & $< 46.51^d$                & $<10.70^d$               & $<3.32^d$              & $<-1.69^d$ \\ 
     &   --      & --           & 2.4$^c$                   & $< 13.28^d$                & $< 42.87^d$                & $<13.42^d$               & $<5.65^d$              & $<-1.60^d$ \\ 
J036.5 & 182.0/156 &   24/  28 & $3.03_{-0.89}^{+1.08}$ & $  7.92_{ -2.19}^{+  2.24}$ & $  2.31_{ -0.65}^{+  0.64}$ & $9.61_{-2.59}^{+2.75}$   & $6.11_{-1.65}^{+1.75}$ & $-1.65_{-0.04}^{+0.05}$  \\ 
J011 &   --      & --           & 2.0$^c$                   & $<  13.78^d$                & $< 49.85^d$                & $<9.40^d$               & $<2.92^d$              & $<-1.54^d$ \\ 
     &   --      & --           & 2.4$^c$                   & $<  13.61^d$                & $< 45.66^d$                & $<11.64^d$               & $<4.90^d$              & $<-1.45^d$ \\ 
J083.8 & 178.0/158 &   53/  42 & $1.89_{-0.37}^{+0.39}$ & $ 13.22_{ -2.47}^{+  2.46}$ & $ 28.99_{ -5.53}^{+  5.35}$ & $11.54_{-2.10}^{+2.19}$  & $3.28_{-0.60}^{+0.62}$ & $-1.66_{-0.03}^{+0.03}$   \\ 
J0050 & 153.4/154 &   31/  35 & $1.89_{-0.40}^{+0.45}$ & $ 19.34_{ -4.01}^{+  4.06}$ & $ 31.22_{ -6.44}^{+  6.38}$ & $11.81_{-2.36}^{+2.49}$  & $3.35_{-0.67}^{+0.71}$ & $-1.70_{-0.03}^{+0.04}$   \\ 
J029 & 192.8/157 &   54/  19 & $2.85_{-0.54}^{+0.60}$ & $ 10.43_{ -2.12}^{+  2.06}$ & $  3.63_{ -0.73}^{+  0.74}$ & $8.25_{-1.63}^{+1.71}$   & $4.71_{-0.93}^{+0.98}$ & $-1.65_{-0.03}^{+0.03}$  \\ 
\hline
J0224 & 180.0/155 &   71/  48 & $2.10_{-0.21}^{+0.22}$ & $ 54.60_{ -6.68}^{+  6.74}$ & $ 57.82_{ -7.08}^{+  7.02}$ & $36.47_{-4.35}^{+4.56}$  & $12.23_{-1.46}^{+1.53}$ & $-1.61_{-0.02}^{+0.02}$  \\ 
J0100 & 146.7/161 &  206/ 156 & $2.39_{-0.12}^{+0.13}$ & $ 69.08_{ -4.33}^{+  4.25}$ & $ 55.40_{ -3.41}^{+  3.45}$ & $57.68_{-3.56}^{+3.67}$  & $24.10_{-1.49}^{+1.53}$ & $-1.72_{-0.01}^{+0.01}$  \\ 

\hline
    \end{tabular}
    \label{analysis}
\tablefoot{$^a$: 0.3-7~keV net-counts; $^b$: MOS1+MOS2 counts; $^c$ fixed parameter, $^d$: 90\% upper limit from pn photometry from Table~\ref{counts} assuming a power-law with the fixed $\Gamma$}
\end{table*}
Given the Type~1 nature of these sources, their high redshift and low number of counts in the spectra we adopted a simple power-law model, i.e. assuming no intrinsic $\nhsym$ for the QSOs,  modified by the absorption by the Galaxy interstellar medium \citep[adopting][maps]{HI4PI2016}, parameterized by a {\em tbabs} model in XSPEC). 
We jointly modelled the three EPIC camera spectra. Given the low-counts regime, we neglect intercalibration shifts between the detectors after checking that they are consistent to unity within the uncertainties.
We performed the fits only for the ten detected sources and carried out the analysis in the energy range 0.3-7~keV (corresponding to rest-frame energies from $\sim2$ to $\sim50$~keV) by leaving free to vary  $\Gamma$ and the normalization. 

The best-fit parameters are reported in Table~\ref{analysis} where the uncertainty on the fluxes and luminosities is computed by freezing $\Gamma$ at its best-fit value. Spectra and best-fit models for the \hyph\ targets are reported in Fig.~\ref{spectra}.
This is the first X-ray spectral analysis reported for the sources J083.8 and J029. Other detected sources were previously observed and analysed with lower quality data, either with \chandra\ and/or  \xmm. A comparison with previous analysis is reported in Appendix~\ref{cfrprevious}.  
Given the background-dominated regime, we verified that changing the  spectral analysis energy range to progressively lower/higher observed energies (i.e. 2, 5, 10~keV) does not significantly impact on our results being always less than 10\% from the best-fit $\Gamma$ and well within the $1\sigma$ uncertainties quoted in Table~\ref{analysis}. 

\subsection{Average spectral slope}\label{jointspec}
To obtain a measure of the average spectral slope from this \hyp\ sub-sample, we performed a joint modeling from all the 10 detected sources. Each QSO dataset, except J0100, contributes to the joint fit with pn+MOS1+MOS2 0.3-7~keV net-counts in the range $\sim50-140$. In the case of J0100, which has more than 300~pn+MOS net-counts, we selected three chunks (chunks1, chunks2 and chunks3) of observations representative of the average pn and MOS net-counts gathered from the other datasets, i.e. with $\sim50\pm20$ and $\sim40\pm15$ net-counts for pn and MOS1+MOS2 detectors, respectively. In order to ensure a random sampling of the observation, the three chunks were selected by adopting a non overlapping count-rate selection of the high-energy light curves used for the high-background screening\footnote{The three chunks were selected by including time intervals of the observation with the  pn/MOS1/MOS2 count-rates ranges 0.19-0.23/0.04-0.06/0.09-0.11 $\rm counts~s^{-1}$ for chunk1, 0.24-0.26/0.065-0.085/0.115-0.135 $\rm counts~s^{-1}$ for chunk2 and 0.27-0.3/0.085-0.115/0.135-0.17 $\rm counts~s^{-1}$ for chunk3.}.
We performed joint pn+MOS1+MOS2 spectral analysis of each chunk and verified that with the simple power-law modeling modified by the Galactic absorption, the $\Gamma$ and the 2-10~keV and 2~keV X-ray luminosities are consistent with those reported for the entire dataset (see Table~\ref{analysis}).

We performed the joint modeling of the 10 QSOs exploiting the 11 datasets (including the two observations of J1342) three times, each analysis including one of the three chunks of the J0100 observation. 
In total we modelled a total of $\sim900$~net-counts of spectral data (0.3-7~keV) of which $\sim500$ and $\sim400$~net-counts are from pn and MOS detectors, respectively.
We adopted a simple power-law model absorbed by Galactic interstellar medium, with $\Gamma$ linked across all the datasets. We included and tied for each detector the cross-calibration constants. We left the linked $\Gamma$ and the normalizations for each source free to vary. A fit to these data resulted in best-fit values of $\Gamma=2.44_{-0.10}^{+0.11}$ (including chunk1 of J0100; $\rm W-stat/dof=1922.8/1730$), $\Gamma=2.40\pm0.11$ (including chunk2; $\rm W-stat/dof=1934.2/1730$) and $\Gamma=2.41_{-0.10}^{+0.11}$ (including chunk3; $\rm W-stat/dof=1944.0/1730$). 

We verified the stability of the results as a function of the energy range and measured that the best-fit value changes within the range $\Gamma=2.39-2.46$ with no trend as a function of energy. Errors on $ \Gamma$ increase from 0.10 to 0.13 by restricting the band interval.

We also removed the datasets with the highest number of net-counts (142~total net-counts; J0244) and lowest net-counts (47 and 52~total net-counts; J036.5 and J0020, respectively)  obtaining substantially unaffected best-fit $\Gamma$ values ($\Gamma=2.37-2.42$). 

Given the good spectral quality (large number of counts) reached in the joint analysis, we also tried to include an intrinsic absorption term to estimate the average hydrogen column density in QSO at EoR. The absorber may be associated with local absorption in the vicinity of the QSO, or with material further out in dense patches of the intergalactic medium. We obtain best-fit $\nhsym$ ranging from $2.1\times10^{21}~\nhi$ and $3.7\times10^{22}~\nhi$ and corresponding slightly steeper $\Gamma$ in the range 2.42-2.62. However the $\nhsym$ are highly uncertain and consistent with no absorption at $\sim1.2-1.3\sigma$ level. Therefore we conclude that mild absorption ($\nhsym\approx10^{21}-10^{22}~\nhi$) is not required in \hyp\ QSOs.

Finally we performed a fit in the same rest-frame energy range for each source. The common energy range is defined as $0.3~\rm keV*(1+z_{\rm max})\,$--$7~\rm keV*(1+z_{\rm min})\approx 2.6-49$~keV, where $z_{\rm min}$ and $z_{\rm max}$ are the highest and lowest redshift covered by the \hyp\ sample considered in this work. We obtained $\Gamma$ ranging from 2.37 to 2.47, with uncertainties of the order of 0.11-0.13. 

We also tried a power-law model with high energy cutoff ($E_{cut}$) under the hypothesis that the steepening of the spectrum is due to the cutoff close or within the relatively high energies covered by the spectral data. By setting $\Gamma$ in the range of canonical values 1.8-2.0 we obtain the energy cutoff $E_{cut}$ in the range $14-25$~keV in all cases. Specifically, the best-fit $E_{cut}$ values for assumed $\Gamma=1.9$ are all in the range $\sim17-19$~keV. Indeed we obtained  $E_{cut}=16.6^{+4.8}_{-3.3}$~keV for J0100 chunk1, $E_{cut}=19.4^{+6.7}_{-4.3}$~keV for chunk2 and $E_{cut}=18.4^{+6.0}_{-3.9}$~keV for chunk3. All the fits are statistically indistinguishable from the simple power-law case having $|\Delta W-stat|\lesssim2$.

\begin{figure}[t]
   \begin{center}
    \includegraphics[width=0.49\textwidth, angle=0]{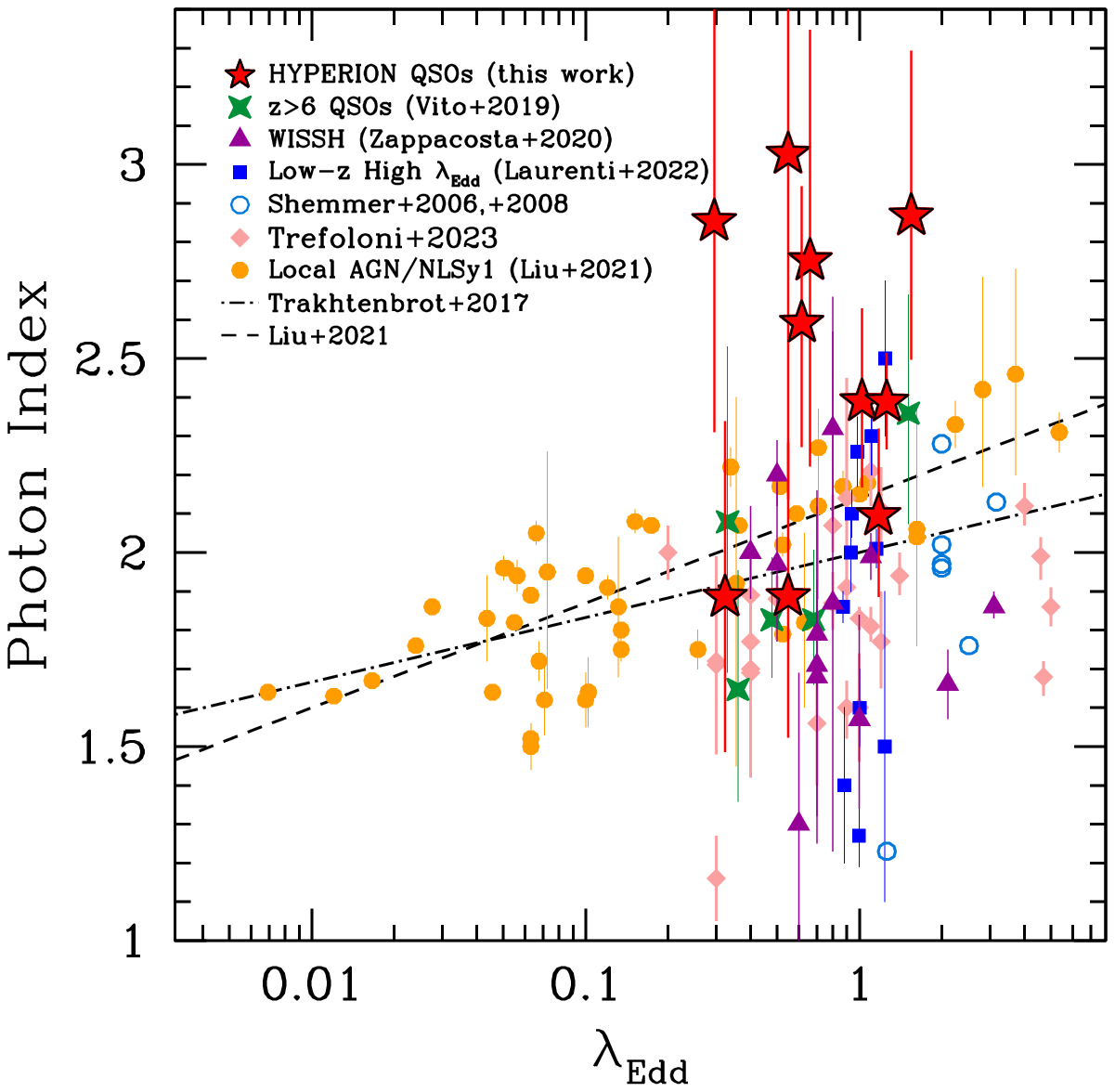}
   \end{center} 
\caption{$\Gamma$~vs~$\lambda_{\rm Edd}$  for a compilation of local/high-$\ledd$ AGN and high-redshift luminous QSOs. Red stars are the \hyp\ QSOs presented in this work. Green four-pointed stars are other $z>6$ QSOs, detected with $\gtrsim30$~net-counts, from the X-ray spectral analysis performed by \citet{vito2019} and not included in the \hyp\ sample. Quasars at Cosmic Noon ($z=2-4$) are reported in purple triangles, \citep[the WISSH QSOs from][]{zappacosta2020},  empty cyan circles \citep{shemmer2006,shemmer2008} and pink diamond \citep{nardini2019,trefoloni2023}. Local high-$\ledd$ QSOs \cite{laurenti2022} and local AGN/NLSy1 \citep{liu2021} are reported respectively as blue square and yellow circles. Reported are also the most recent relations from a linear fit to the local AGNs \citep[i.e.][]{trakhtenbrot2017,liu2021}. The uncertainties on $\ledd$ from QSOs with $\mbh$ estimated by single-epoch virial mass estimator are dominated by systematic uncertainties and can be as high as 0.5~dex. The statistical uncertainty on $\ledd$ for the local AGN/NLSy1 \citep{liu2021} whose masses are estimated via reverberation mapping is 0.1~dex and 0.2~dex for the sub-Eddington and super-Eddington sources, respectively.}
       \label{gammaledd}
\end{figure}

\begin{figure}[t]
   \begin{center}
    \includegraphics[width=0.49\textwidth, angle=0]{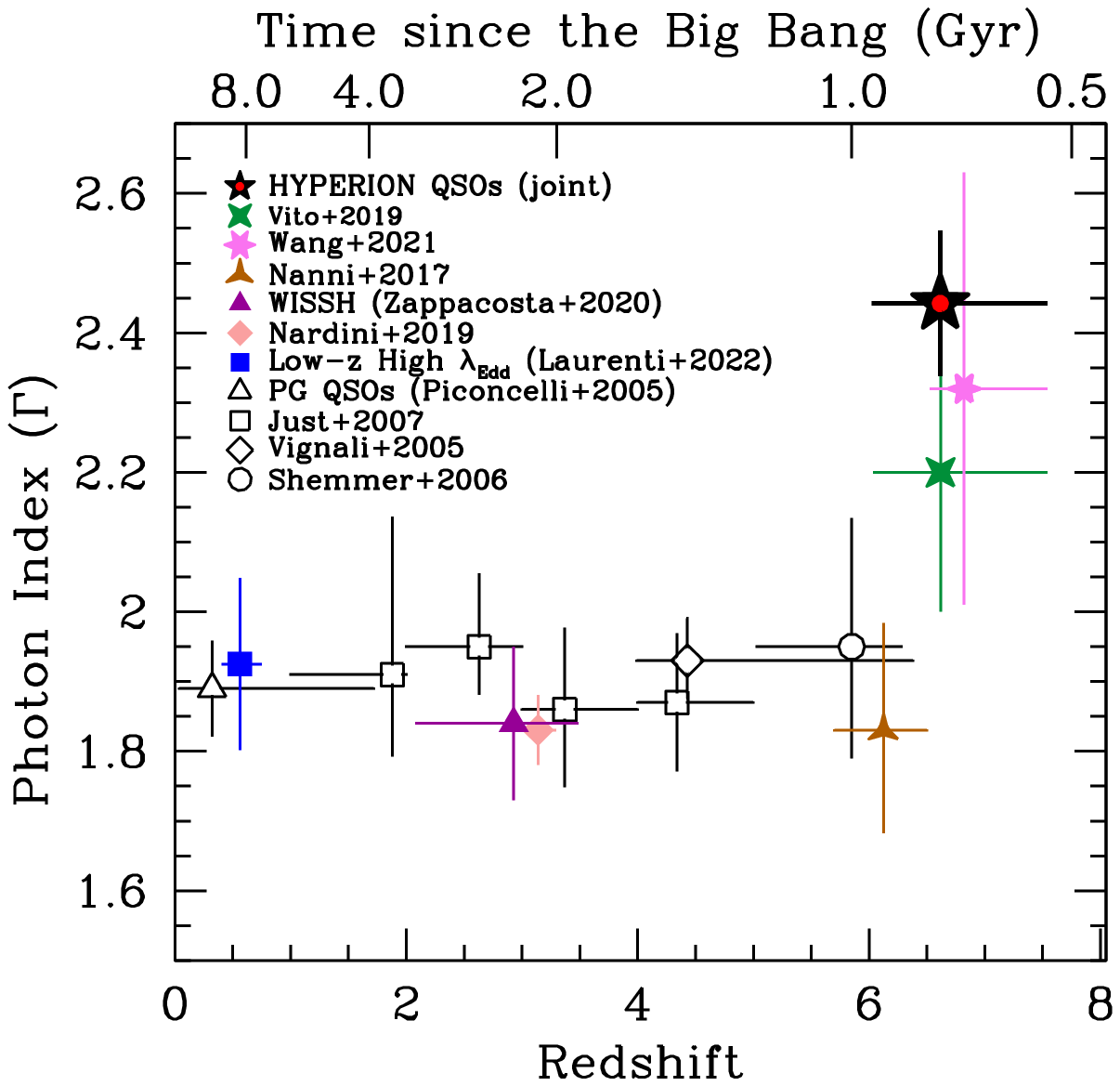}
   \end{center} 
\caption{Distribution of the average $\Gamma$ as a function of redshift. The plot includes data from joint spectral analysis or average values from samples of QSOs. In particular, starred data are from joint spectral analysis of samples of $z>6$ QSOs. Black star with central red circle, green four-pointed star, magenta six-pointed star and three-pointed star are \hyp\ QSOs, \citet{vito2019}, \citet{wang2021} and \citet{nanni2017} samples, respectively. The empty squares, diamond and circle are averages from stacked spectral analysis of luminous and hyperluminous QSOs from  \citet{just2007}, \citet{vignali2005} and \citet{shemmer2006b}, respectively. The empty triangle represent the average $\Gamma$ from the PG quasars \citep{piconcelli2005}. Blue squares are high-$\ledd$ local QSOs \citep{laurenti2022} and purple triangles are hyperluminous high-$\ledd$ WISSH QSOs \citep{zappacosta2020}. Pink diamonds are $z\sim3$ luminous blue quasars from \citet{nardini2019}. Vertical error bars report $1\sigma$ uncertainties on $\Gamma$  while horizontal error bars indicate the redshift range covered by the QSO sample considered in each dataset.}
       \label{gammaz}
\end{figure}

\begin{figure}[t]
   \begin{center}
    \includegraphics[width=0.49\textwidth, angle=0]{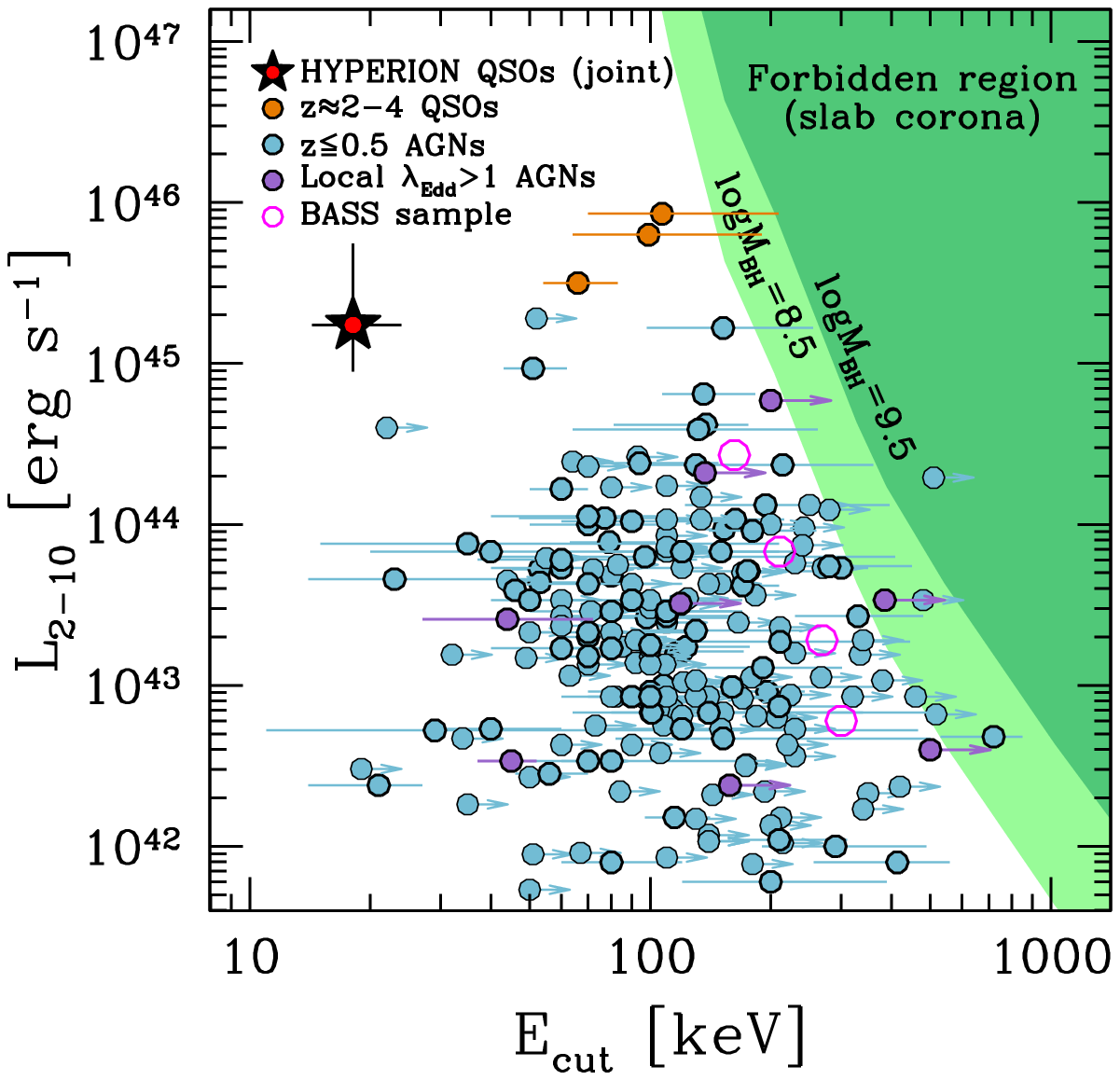}
   \end{center} 
\caption{Distribution of $E_{cut}$ as a function of $\lumh$. Light blue and orange filled circles are estimates from a compilation of local AGN \citep[][and references therein]{bertola2022} and $z\approx2-4$ QSOs \citep{lanzuisi2019,bertola2022}. Purple circles are from local super Eddington accreting AGN from \citet{tortosa2023}. Hollow magenta circles are binned average estimated from \citep{ricci2018} of a large sample of local AGN from the BAT AGN Spectroscopic Survey (BASS). \hyp\ average $E_{cut}$ measurement (assuming $\Gamma=1.9$) from our joint analysis is reported as black star with inner red circle. Green regions are the forbidden regions (for a slab corona model) due to runaway electron-positron pair production \citep[see][]{svensson1984} for $\log (\mbh/\msun)=8.5$ and $\log (\mbh/\msun)=9.5$. 
}
       \label{ecutlx}
\end{figure}

\begin{figure*}
   \begin{center}
    \includegraphics[width=0.49\textwidth, angle=0]{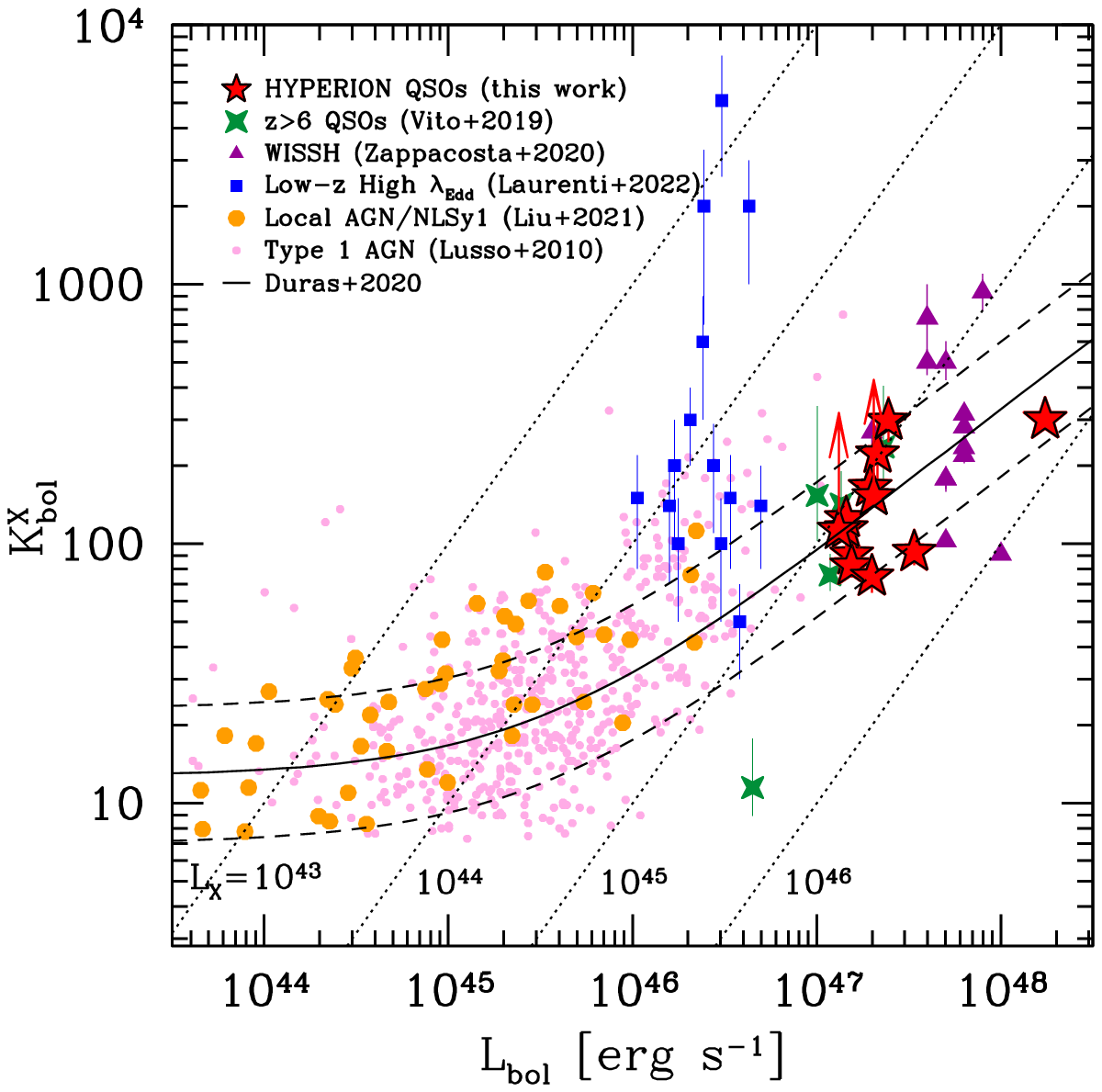}
    \includegraphics[width=0.49\textwidth, angle=0]{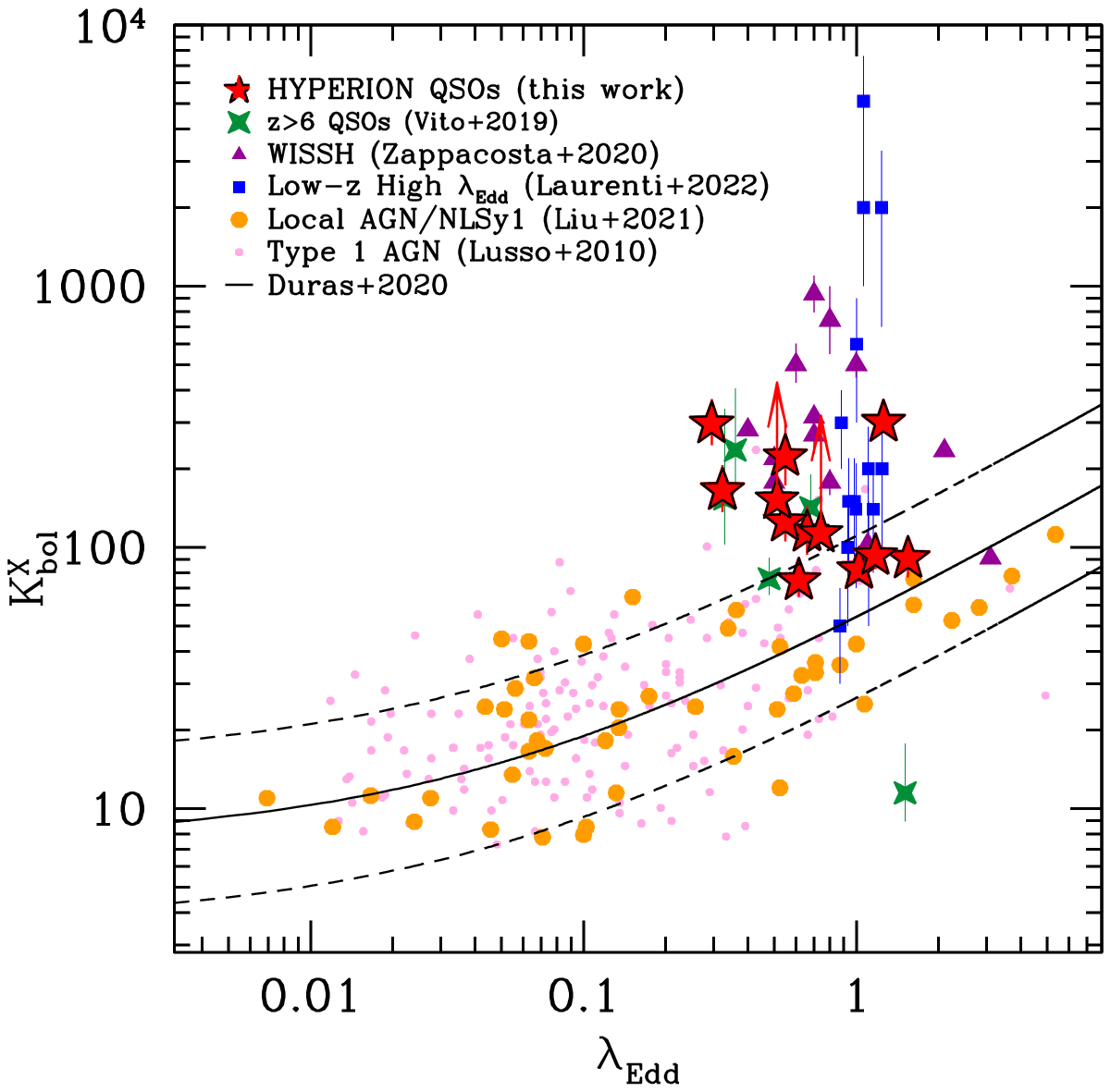}
   \end{center} 
\caption{Bolometric X-ray luminosity as a function of $\lbol$ and $\ledd$. Left: $\kbol$~vs.~$\lbol$ for a compilation of broad-line mostly high-z QSOs and local AGN. The sources are reported as in Fig.~\ref{gammaledd}. Lower limits for the \hyp\ QSOs are estimated assuming a power-law with fixed $\Gamma=2.4$ (see Section~\ref{counts} and Table~\ref{specanalysis}). We also added COSMOS Type~1 AGN (pink dots) from \citet{lusso2010}. Solid and dashed black lines represent the fitting relation reported in \citet{duras2020} and its $1\sigma$ spread. Dotted lines report fixed value of 2-10~keV  luminosity in units of $\lumcgs$ in the $\kbol$~vs.~$\lbol$ plane.
Right: $\kbol$~vs~$\ledd$ for the same sources reported in left panel and with SMBH measures available. Solid and dotted black lines represent the fitting relation reported in \citet{duras2020} and its $1\sigma$ spread.
}
       \label{kbol}
\end{figure*}

\begin{figure}
   \begin{center}
    \includegraphics[width=0.49\textwidth, angle=0]{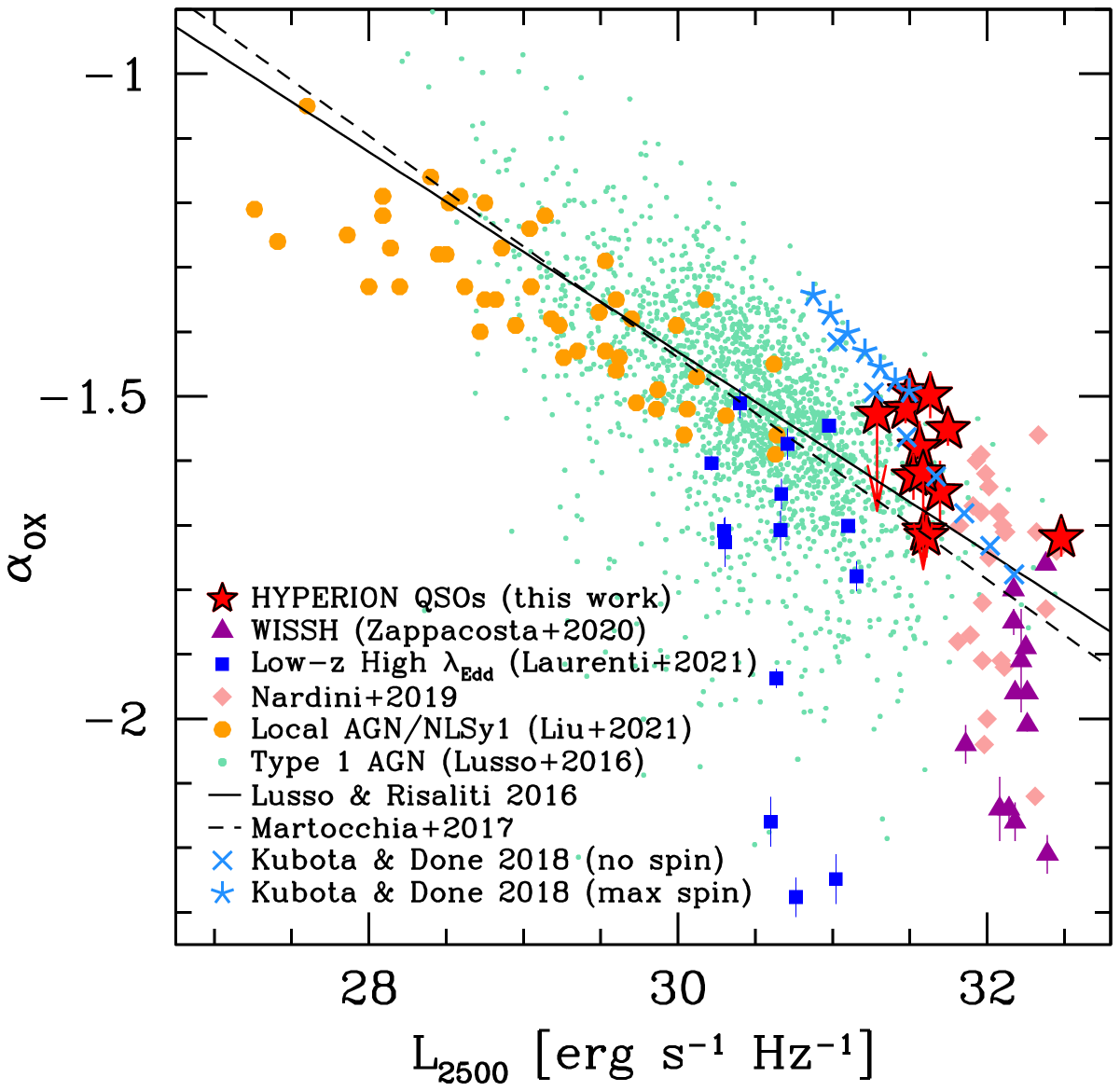}
   \end{center} 
\caption{$\aox$~vs~$\luvmon$ for a compilation of AGN catalogs. Symbols are referred to the AGN samples as in Fig.~\ref{gammaledd} except for the green dots which are detected AGN with signal-to-noise ratio $SNR> 5$ from \citet{lusso2016}. Upper limits for the \hyp\ QSOs are estimated assuming a power-law with fixed $\Gamma=2.4$ (see Section~\ref{counts} and Table~\ref{specanalysis}).  Dashed line is the linear fit from \citet{martocchia2017} while solid line refers to the best-fit relation from \citet{lusso2016} and for the sub-sample with $SNR>5$, E(B-V)$\geq0.1$ and $1.6\leq\Gamma_{1-5}\leq2.8$, $\Gamma_{1-5}$ being the photon index estimated between the luminosities at  1~keV and 5~keV. Light blue crosses and asterisks present the values predicted by the {\tt QSOSED} model \citep{kubota2018} assuming average \hyp\ parameters and spin $a=0$ and $a=1$, respectively. They are reported from top-left to bottom-right from $\log \dot{m}=-1$ to $\log \dot{m}=0.2$ in steps of $\Delta \log \dot{m}=0.2$.}
       \label{aoxluv}
\end{figure}

\begin{figure}
   \begin{center}
    \includegraphics[width=0.49\textwidth, angle=0]{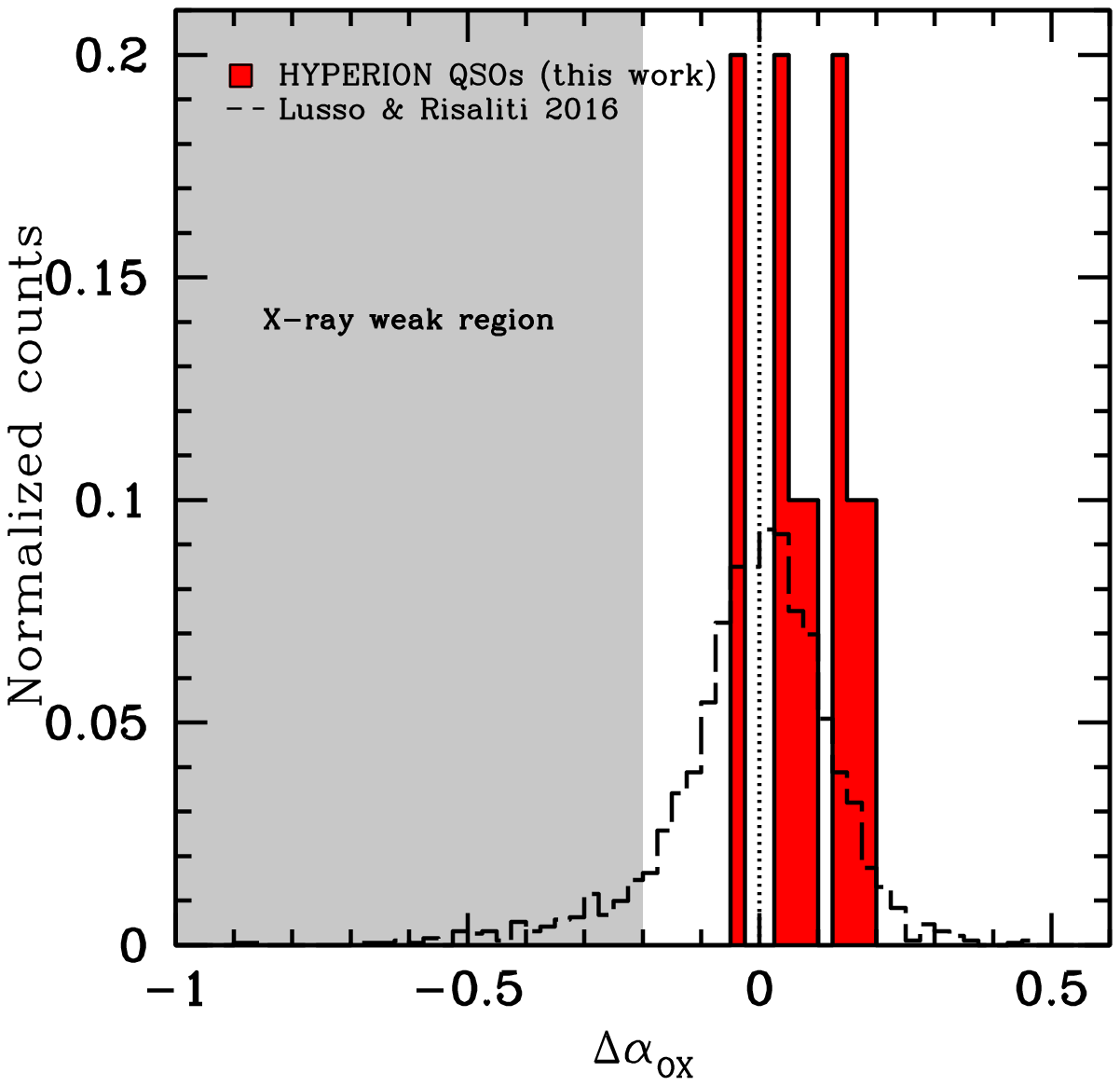}
   \end{center} 
\caption{Normalized distributions of the $\Delta\aox$ relative to the relation from \citet{lusso2016} (see Fig.~\ref{aoxluv}) for \hyp\ QSOs (red filled histogram) and the AGN detected with $SNR>5$ from \citet{lusso2016} (dashed hollow histogram). The grey region marks the position of the X-ray weak sources.}
       \label{daoxhisto}
\end{figure}

Notice that in our joint and single source analysis we neglected  contributions from a Compton reflection component due to the coronal X-rays inverse Compton-scattered by the surrounding matter. Typically QSOs show low or virtually no reflection \citep[e.g.][]{vignali1999,reeves2000,page2005,zappacosta2018}. A non-negligible Compton reflection contribution in the \hyp\ QSOs would result in even steeper $\Gamma$ for their power-law continuum.

\section{Results}\label{results}
 In the following, we compare the X-ray properties inferred from our analysis of the \hyp\ spectra with those reported for other $z\leq6$ sources, especially luminosity- and $\ledd$- analogs QSOs to assess possible differences linked to radiative output or accretion rate as parameterized by the $\ledd$ or to the SMBH mass accretion history stage of the SMBH adopting as a proxy $\mseed$. 
\subsection{The steepness of the X-ray spectrum}\label{discussionsteepness}
The $\Gamma$ measured for each \hyp\ QSO are on average very steep. Fig.~\ref{gammaledd} shows the distribution of $\Gamma$ as a function of $\ledd$ for our \hyp\ QSOs and other AGN and QSOs. Recent relations measured for local sample of low-luminosity AGNs involving these two quantities are reported to aid with the interpretation of the plot \citep{trakhtenbrot2017,liu2021}. \hyp\ sources show on average the steepest values, the large majority exhibiting $\Gamma\geq2.3$. Other $z>6$ QSOs with good quality data (i.e. $>30$ net-counts) not included in \hyp\ (i.e. which failed the $\mseed$ selection criterion) from \citet{vito2019} show flatter $\Gamma$ values. Other $\ledd$-analog QSO samples at lower redshift, i.e. the hyperluminous WISSH $z=2-3$ QSOs from \citet{zappacosta2020} and the high-$\ledd$ nearby ($z<1$) QSOs from \citet[][hereafter L22]{laurenti2022}, have noticeably flatter $\Gamma$ (although with large scatter) in agreement with the canonical $\Gamma=1.8-2$ values. The reported relations predict $\Gamma\approx1.9-2.1$ at the average \hyp\ $\ledd$. The average $\Gamma$ for \hyp\ QSOs from the joint spectral analysis is inconsistent with the relations at $>3\sigma$ level.  

 In the $\Gamma$ vs. $z$ plot reported in Fig.~\ref{gammaz} we show our joint analysis $\Gamma$ value compared with the results of other, independent, joint analysis studies of $z\gtrsim6$ QSOs \citep[][]{nanni2017,vito2019,wang2021}.
 Thanks to the combination of a sizable number of detected sources with the higher quality data gathered by the 1st year of \hyph, the uncertainty in our average joint value is smaller by a factor 2-3. 

We also report in the plot previous joint spectral analysis $\Gamma$ values from other luminous QSO sample at $1<z<6$  \citep{vignali2005,shemmer2006,just2007}, the average value for the WISSH QSOs from \citet{zappacosta2020}, for the local PG QSOs \citep{piconcelli2005} and the high-$\ledd$ L22 QSOs. These are  samples of analog sources in terms of $\lbol$ and/or $\ledd$. All $z<6$ results from $\lbol$- and $\ledd$- analogs sources show consistency with $\Gamma=1.8-2$. The average $\Gamma$ from all the considered $z<6$ QSO samples is $\Gamma_{z<6}=1.91\pm0.04$. 

\hyp\ QSOs have a  $\Gamma$ value inconsistent with $\Gamma=2$ at $>4\sigma$. The same inconsistency level holds with the $\Gamma$ reported for $z<6$ sources of similar $\lbol$ or $\ledd$. In particular $\Gamma$ for \hyp\ is inconsistent at $\sim4.8\sigma$ level with $\Gamma_{z<6}$. We also measured for the WISSH QSO sample analyzed by \citet{zappacosta2020}, the average $\Gamma$ obtained by performing the spectral fits from 2~keV, i.e. the same rest-frame low-energy bound probed for the \hyp\ QSOs. We obtained an average $\Gamma=1.93\pm0.08$ which is consistent with the average value of $\Gamma=1.84\pm0.07$ inferred from the full bands (i.e. from 0.2-0.3~keV observed frame low energy bound, corresponding to 0.6-0.9~keV rest-frame) spectral modelings. This further indicates that the steepness of the \hyp\ $\Gamma$ values does not depend on the probed rest-frame energy range.
Consistency between the \hyp\ $\Gamma$ value and those from past works analysing $z>6$ samples is reported at the $1-2\sigma$ level. This is due to the large uncertainties reported in past $z>6$ QSO analysis.

All the previous comparisons  suggest that the $\Gamma$ of \hyp\ QSOs is steeper regardless of the QSOs luminosity or accretion rate and hence that it is due to an evolutionary effect. Given the \hyp\ QSOs selection criterion this evolutionary effect is  possibly linked to the particularly fast SMBH mass growth history undergone by these sources.  

In order to check this hypothesis, we divided the 10 \hyp\ QSOs in two equal size samples according to their SMBH growth history and hence based on their required $\mseed$. Specifically we selected: (i) a high $\mseed$ sample (i.e. $\mseed\approx4-30\times10^{3}~\msun$; including J1342, J1120, J0100, J0020, J036.5) and (ii) a low $\mseed$ sample (i.e. $\mseed\approx1-3\times10^{3}~\msun$; including J0244, J0224, J0050, J083.8, J029). X-ray data for each sample include an approximately equal number of pn+MOS1+MOS2 counts, with 410 and 495 net-counts for the high-$\mseed$ and low-$\mseed$ samples, respectively. We performed a joint spectral analysis for each sample. For the high-$\mseed$ sample we obtained $\Gamma=2.64-2.7$ (depending to the J0100 chunk used), with uncertainty of $\sim0.16$,  and for the low-$\mseed$ sample we obtained $\Gamma=2.21\pm0.13$, a difference that is significant at the $2.1-2.4\sigma$ level. The average redshift of each sample is $6.86$ and $6.37$ for the high-$\mseed$ and low-$\mseed$ samples, respectively. Therefore the $\Gamma$ difference can be also due to a redshift (i.e. temporal) dependence. Indeed, redshift and $\mseed$ in this sample correlate with a Spearman rank correlation coefficient of $\sim0.8$. This is probably due to Malmquist bias as the virial mass estimators are luminosity dependent.
A joint analysis on the five lowest/highest redshift QSOs ($z=6.29$ and $z=6.94$, respectively) gives  $\Gamma=2.21-2.29$ and $\Gamma=2.64^{+0.17}_{-0.16}$ with $\sim0.14$ uncertainties, confirming the increasing $\Gamma$ trend with redshift. However, this $\Gamma$ steepening, if confirmed, is happening in $\sim10^8$~years which is a very short period of time for a any likely redshift-dependent mechanism to act on cosmological timescales. Hence, we support the hypothesis that the steepening (if confirmed by additional data) is dependent on $\mseed$ and hence on the rapid mass growth of the SMBH.

A steep spectrum can also be mimicked by a power-law with canonical $\Gamma=1.9$ and high-energy cutoff at relatively low energies. Our data are not able to discriminate between a simple power-law or a cutoff power-law model, hence we cannot rule out this possibility. Fig.~\ref{ecutlx} shows the distribution of energy cutoff $E_{cut}$ as a function of $\lumh$. The \hyp\ QSOs considered in this work are compared to $z<0.5$ lower-luminosity AGN and to $z=2-4$ hyperluminous lensed QSOs \citep{lanzuisi2019,bertola2022} as well as to local super-Eddington accreting AGN from \citet{tortosa2023}. The \hyp\ value of $E_{cut}$ is at extremely low energies and, although consistent with few measurements for low-luminosity AGN, is inconsistent with the few measurements for QSOs at similar $\lumh$. Furthermore,  it is far from the forbidden area in which runaway electron-positron pair production would act as thermostat lowering the temperature of the corona and hence $E_{cut}$ \citep[see][]{svensson1984,stern1995}. The forbidden region extent is $\mbh$ dependent and is calculated from \citet{fabian2015} assuming a slab geometry (we do not show the, less extended and hence less conservative, hemisphere geometry regions).

\citet{ricci2018} found a statistically significant anti-correlation between $E_{cut}$ and $\ledd$ for a sample of local AGN from the BASS survey \citep{koss2017} and up to $\ledd\approx0.4$. An extrapolation of this relation to the average $\ledd=0.8$ (or $\ledd=0.5$ if adopting the mass estimator from \citealt{shen2011}) of the \hyp\ sub-sample studied in this work indicates values as low as $100$~keV (i.e. accounting for the uncertainty given by the median absolute deviation of this relation). Our $E_{cut}$ is inconsistent at $>3\sigma$ level with the trend of this relation (i.e. the $3\sigma$ upper bound is $\sim60$~keV).

\subsection{Comparing the X-ray contribution to the UV/bolometric radiative output}
We also check the behaviour of the X-ray coronal luminosity of the \hyp\ QSOs to the bolometric radiative output. 
The bolometric correction $\kbol=\lbol/\lumh$ has a somewhat flat trend at Seyfert-like luminosities progressively increasing toward higher luminosity sources \citep[e.g.][]{marconi2004,lusso2012,duras2020}. 
Fig.~\ref{kbol} left, shows the bolometric correction $\kbol$ as a function of $\lbol$. The \hyp\ QSOs are in agreement with the trend delineated by other data (except the nearby optically-selected high-$\ledd$ QSO from L22) and described by the relation of \citet{duras2020}. Despite this, the location of the \hyp\ QSOs in the $\ledd-\kbol$ plane as reported in the right panel of Fig.~\ref{kbol}, appears to be in disagreement with the trend reported by \citet{duras2020}.  This disagreement is shared by all QSO samples and highlights the lack of a clear dependence between $\kbol$ and $\ledd$. 
This is mainly due by the steep $\lbol$ dependence of $\kbol$ at high-luminosity regimes\footnote{The only exception being the lower luminosity L22 sample which mainly deviates because of their overall X-ray weakness, possibly a result of a optical selection coupled to the high-$\ledd$ requirement.}. This is not well sampled by \citet{duras2020} and is dominated by the bulk of low-luminosity highly accreting AGN population. 

We now investigate the $\aox$ parametrizing the slope between the monochromatic luminosities at 2~keV and 2500~$\rm\AA$ and defined as $\aox=\log (\lxmon/\luvmon)/\log (\nu_{\rm 2~keV}/\nu_{2500\rm ~\AA})$. 
Fig.~\ref{aoxluv} report $\aox$ vs. $\luvmon$ for several AGN samples spanning $>4$ decades in $\luvmon$ along with best-fit relations from \citet{lusso2016} and \citet{martocchia2017}. 
Unlike other hyperluminous or high-$\ledd$ QSOs exhibiting, on average, a weaker X-ray emission compared to the UV one, the \hyp\ QSOs exhibit, on average, slightly higher 2~keV  luminosities which almost systematically exceeds the expectation of the $\aox$~vs~$\luvmon$ relations with no sources exhibiting the X-ray weakness typically shown by a consistent fraction of sources in the WISSH and L22 samples. Indeed, at the mean $\log \luvmon$, the \citet{lusso2016} relation, providing the more accurate parameterization of the bulk of the AGN population, predict $\aox=-1.69$ while the average for the detected \hyp\ QSOs has $\aox^{\scriptstyle hyp}=-1.61\pm0.030$. This translates to an average $\Delta\aox^{\scriptstyle hyp}=0.08$.  
We computed the distribution of the $\Delta\aox$ values for \hyp\ and the \citet{lusso2016} QSOs detected with a signal-to-noise ratio $SNR>5$ (see Fig.~\ref{daoxhisto}) and performed a Kolmogorov-Smirnov test on the detected data  to check the difference between the two datasets. The two distributions mildly differ with a null-hypothesis probability $P_{null}=0.0576$. We further check this  disagreement performing 10000 random draws of $\Delta\aox$ in sub-samples of 10 sources (i.e. the same size of the detected \hyp\ QSO sample reported in this work) from the \citet{lusso2016} sample. We verified that their average $\Delta\aox$ do not exceed that shown by the \hyp\ QSOs for 98.4\% of the time and, therefore, that the disagreement of our sources is not strong. 
This result is also slightly at variance with the $\aox$ previously estimated for $z>6$ QSOs by other works \citep{vito2019,wang2021}. This is mainly due to the fact that they assumed a $\Gamma=2$ to derive $\lxmon$. 
Hence, we can partly explain this as the combination of steep $\Gamma$ and unchanged integrated $\lumh$, (i.e. in line with the values expected by the $\kbol$~vs~$\lbol$ relation). Indeed, at fixed $\lumh$, a change in $\Gamma$ slope from 2 to 2.4 increases $\lxmon$ by a factor $\sim1.3$ and hence $\aox$ by a quantity $\sim0.11$, in agreement with $\Delta\aox^{\scriptstyle hyp}$. 

Assuming a best-fit power-law model with a high-energy cutoff and canonical $\Gamma=1.9$, would result in average $\aox=-1.65\pm0.071$, i.e. somewhat softer and more in line with the $\aox$~vs~$\luvmon$ relations, hence exhibiting a $\Delta\aox=0.042$.

Notice that at this high luminosity regime there is a clear contrast between the \hyp\ QSOs and the QSOs at Cosmic Noon ($z=2-4$). Indeed, the latter QSOs are characterized by flatter slopes (on average $\Gamma\approx1.85$) translating to $\aox$ smaller by a quantity $\sim0.2$ than those shown by the former QSOs. Furthermore a fraction corresponding to $\sim30$\% of the WISSH QSOs and the luminous blue QSOs at $z=3$ analyzed by \citet{zappacosta2020} and \citet{nardini2019} are characterized by intrinsic X-ray weakness further lowering their average $\aox$ values.

\section{Probing a new regime in the nuclear properties of QSOs at EOR}\label{discussion}
The measured X-ray properties of \hyp\ QSOs clearly differ from  luminosity-analog and $\ledd$-analog QSOs at lower z. Steep X-ray spectral slopes (regardless of wheter they are due to steep $\Gamma$ or to a low-energy onset of the power-law cutoff) as  measured here are previously unreported among the QSO population. They are more typical of lower $\mbh$ ($<10^6~\rm \msun$) highly accreting low-luminosity AGN such as the narrow line Sy1  (NLSy1) galaxies \citep[e.g.][]{miniutti2009,ludlam2015}. The steep $\Gamma$ measured for the \hyp\ QSOs are also a confirmation of results reported on single but peculiar $z>6$ sources such as the very bright radio loud \citep{medvedev2021} or narrow-line quasars \citep{wolf2023} for which $\Gamma=2.5\pm0.2$  (90\% errors) and $\Gamma=3.2^{+0.7}_{-0.6}$ have been obtained. 

It is possible that the steepness derives from a different geometry of the accretion disc/corona system, a different coupling between the accretion disc and the corona, or to peculiar coronal properties. 

\citet{kubota2018} present a framework of a radially stratified accretion disc with standard outer disc, inner warm Comptonizing and a innermost hot corona regions,  adopting a truncated disc geometry with  the corona dissipating power in the inner hot accretion flow \citep[see Fig.~2 in][]{kubota2018}. By imposing in the model a fixed $0.02 \lumedd$ fractional dissipation from the hot flow they are able to obtain an increasing hard X-ray $\Gamma$ dependence on the Eddington-normalized accretion rate ($\dot{m}=\dot{M}/\dot{M}_{\rm Edd}$, where $\dot{M}$ is the mass accretion rate and $\dot{M}_{\rm Edd}=\lumedd/c^2$). The relation they found is somewhat steeper than the most recent $\Gamma$~vs~$\ledd$ relations reported in Fig.~\ref{gammaledd}. Their model is however in broad agreement with the measured $\aox$~vs~$\luvmon$ relations. In Fig.~\ref{aoxluv} we show the prediction of their {\tt QSOSED} model (a simplified variant of their model with assumptions tuned for QSOs) for a non rotating (i.e. spin parameter $a=0$) and maximally rotating ($a=0.998$) SMBH,  for different $\log \dot{m}$ from -1 to 0.2 and adopting the average $\log (\mbh/\msun)=9.43$ value for the \hyp\ QSOs. For each $\log \dot{m}$ step, we normalized the model to the \hyp\ average $\lumh=1.7\times10^{45}~\rm \lumcgs$.

\begin{figure}[!t]
   \begin{center}
    \includegraphics[width=0.49\textwidth, angle=0]{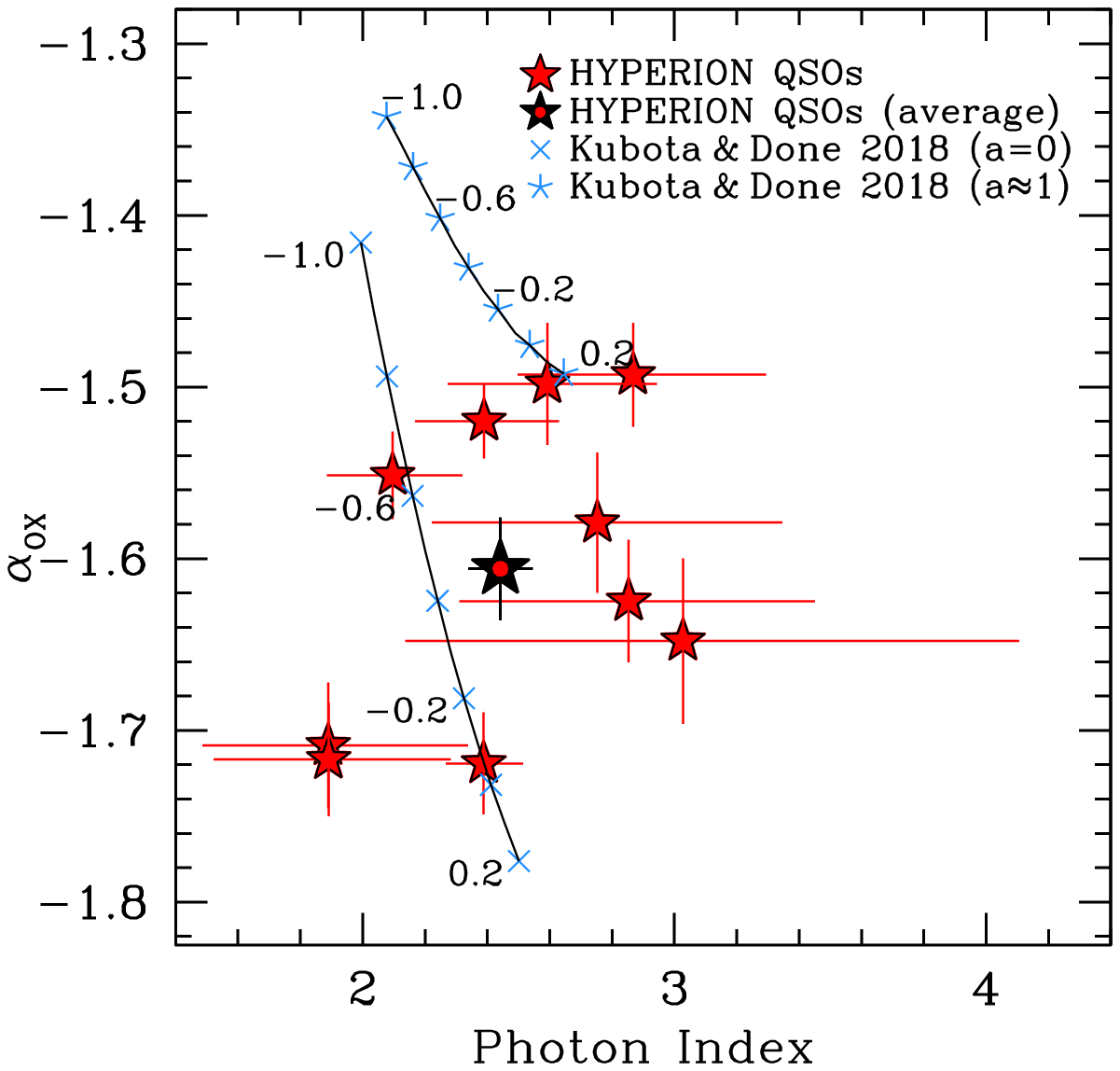}
   \end{center} 
\caption{$\Gamma$~vs~$\aox$ for \hyp\ QSOs compared to the {\tt QSOSED} model prediction. Red stars are single \hyp\ QSOs, while the black star with central red circle show the average $\aox$ and joint best-fit $\Gamma$ for the detected \hyp\ QSOs. Light blue crosses and asterisks are the model predictions (for $a=0$ and $a\approx1$, respectively) whose label report the $\log \dot{m}$ value.}
       \label{gammaaox}
\end{figure}

The $\aox$ predicted by {\tt QSOSED} for non rotating SMBH is consistent with the $\aox$ measured for the \hyp\ QSOs. A maximally spinning SMBH case disagrees with the data predicting flatter $\aox$. Hence, the preference for the null spin case would suggest a scenario of chaotic SMBH accretion flows rather than a more regular secular flow of accretion from the galaxy disc which would instead lead to a conservation of the angular momentum.  
Fig.~\ref{gammaaox} shows the $\Gamma$~vs~$\aox$ plot for the \hyp\ QSOs and their joint/average value compared to the {\tt QSOSED} model predictions. Predicted values are in good agreement with the QSO data suggesting for the \hyp\ QSOs an average $\log \dot{m}\approx-0.4$ (which is close to the average $\Gamma$ and $\aox$ for the sample analysed here) for the non-spinning BH case.

In Section~\ref{discussionsteepness} we report a marginal indication that $\Gamma$ is further steepening at higher redshifts and/or for the sources requiring for their SMBH formation the highest $\mseed$. 
In the framework of the \citet{kubota2018} model and in general taking as a reference the anticorrelation trend of $\Gamma$~vs~$\ledd$, a redshift dependence would translate in higher accretion rate, meaning that the highest redshift sources are still accreting, on average, at high $\dot{m}$ compared to the bin with measured flatter $\Gamma$. The more likely $\mseed$ sub-sample division case points to a scenario in which the fastest SMBH mass accretion history are still highly accreting compared to those which had slower accretion pathways and would probably result on average in future even more massive SMBH. 

\subsection{On the origin of the steep X-ray spectral slopes}
We can parameterize the steep spectral slope also with a power-law with canonical $\Gamma=1.9$ and a low high-energy cutoff. We measured $E_{cut}\approx20$~keV which is very low compared to the currently measured values reported in Fig.~\ref{ecutlx} but well constrained as it falls within the rest-frame energy band.  We are not able to discriminate between a simple power-law model and one with the addition of a low-energy cutoff. In this regard it is worth to mention that the high quality spectrum (i.e. 1400~net-counts at 0.2-10 keV) analyzed by \citet{medvedev2021} for the $z=6.18$ radio loud QSO CFHQS~J142952+544717 does not show any signature of such a low-energy cutoff, providing a lower limit at 30-50~keV, according to the different model parameterization (although a possible X-ray jet component may impact the cutoff  detectability). Since our result applies to a well defined population of QSOs,   it is worth discussing the possibility that also a low-energy cutoff could have originated the reported steep spectral slopes. 
Interestingly, a few local highly accreting Seyfert~1 galaxies with low energy cutoff similar (i.e. 20-30~keV) to those measured in \hyp\ QSOs have already been reported  \citep{vasudevan2013,kamraj2018}.

These values corresponds to coronal temperatures as low as $\sim7-10$~keV. 
There are a number of possible explanations to account for such low temperature coronae. They involve either the interaction (coupling) between a highly accreting accretion disc and the corona or peculiar physical states of the corona. 
 In super-Eddington sources, such as the NLSy1, the strong soft disc radiation field is capable to effectively increase the Compton cooling of the corona, leading to a steep spectrum with a low-energy cutoff \citep{pound1995}. Recent JWST observations support the hypothesis of super-Eddington accreting AGNs in the early universe \citep[$z>8$; e.g.][]{larson2023,maiolino2023}. In this case we can expect for them similar X-ray nuclear properties (i.e. steep photon index and/or relatively low-energy cutoff). 

 Radiatively-driven winds launched from the accretion disc of highly-accreting SMBH can provide an alternative explanation for the steep spectral slopes. If these winds had mass ejection rates larger than the disc mass accretion rates, it is possible that they carry away matter from the innermost disc regions at a rate higher than the mass accretion rate effectively causing the truncation of the disk well before the innermost stable orbit. This would force the corona to be irradiated and hence Compton-cooled by softer seed photons  \citep{laor2014,kara2017}. 
 A recent result reported by \citet{bischetti2022}, in  the XQR-30 sample \citep[][]{dodorico2023} of bright $z\sim6$ QSOs, show a very high fraction of broad-absorption line winds, i.e. a factor 2.4 larger than in low-z QSOs, with velocities up to 17\% the speed of light. This may lend support to this scenario accounting also for a redshift evolution \citep[see also][]{bischetti2023}. 
 So far only two \hyp\ QSOs have been reported to host a BAL \citep[i.e. J0038 and J231.6, ][]{wang2018,bischetti2022}. This indicates a BAL fraction of $11^{+14}_{-7}$\%, where the uncertainties accounts for Poisson statistics, to be compared to the $47^{+16}_{-12}$\% reported in the XQR-30 sample. Assuming a 47\% BAL fraction, the probability of having only two BAL in a sample of 18 QSOs (as in \hyp) by chance is 0.2\%. Notice, however, that homogeneous high-quality spectroscopic datasets  (i.e. SNR$\gtrsim15$, $R\sim6000$) for all \hyp\ QSOs are not available. Hence, a dedicated study to compare  the occurrence of BAL in the \hyp\ and XQR-30 samples, at the same sensitivity level is currently not possible. In any case, a non detection of a large BAL fraction is per se not an indication of lack of nuclear winds. CVI emission line blueshifts (relative to MgII) have been measured for a large part of the \hyp\ QSOs in several works indicating fast broad-line winds up to $5000-6000~\rm \kms$ \citep[e.g.][]{mazzucchelli2017,meyer2019,shen2019,schindler2020,yang2021}. 

A similar disc-truncation scenario may also occur if the inner disc regions are impacted by tidal disruption-like events (TDEs) which increase the accretion rate of the inner regions as a consequence of fast angular momentum removal of the perturbed accreting material by bound debris streams \citep{chan2019} as recently suggested for the reported changing-look event in the low-luminosity local AGN 1ES~1927+654 \citep[][]{ricci2020,masterson2022}. Such events, which in the most extreme cases may lead to the corona destruction as the magnetic field pattern powering the corona is suppressed, are expected to show a X-ray softer-when-brighter behaviour \citep{sobolewska2009}. This cannot be explored with this dataset given the large uncertainties in $\Gamma$ for each source. It is interesting though to notice that in the re-brightening phase of 1ES~1927+654, the X-ray spectrum appears very soft (i.e. with a steep $\Gamma\gtrsim3$) like some of our QSOs and with an additionally low $E_{cut}<20~\rm keV$. In order for this scenario to be applicable to our $z>6$ QSOs these events needs to be frequently recurring so that statistically the sources are caught on average with a soft spectrum. Optimistic TDE rates for $10^9-10^{10}\rm~\msun$ SMBHs are in the range $\sim5\times10^{-6}-10^{-5}\rm~yr^{-1}$ \citep{stone2016}. This implies a TDE event every $1-2\times10^5\rm~yr$. \citet{chan2019} speculates that it will take at most decades before the disk and therefore the corona return in the unperturbed state. This timescale is orders of magnitudes shorter than the TDE timescale and makes this scenario unfeasible.

Alternatively, peculiar corona properties may result in lower temperatures. 
In high optical depth coronae, disc seed photons may undergo multiple scatterings before leaving the corona, hence effectively lowering its temperature \citep{tortosa2017}. 
Finally in hybrid coronae models \citep{zdziarski1993,fabian2017} in which thermal and non thermal particles coexist in a highly magnetized plasma, the heating and cooling processes are faster than the electron cooling time by inverse Compton. In this case even a small fraction of non thermal electrons having MeV energies can cause intense runaway electron-positron pair-production. The cooled pairs may afterward redistribute their energy to the thermal particles therefore lowering the temperature of the corona. Although attractive, for these scenarios still a redshift dependence justification cannot be easy to justify. 

Finally, an interesting scenario may couple the presence of optically thick coronae to the occurrence of nuclear winds which, as we already discussed, may provide a justification for a $z>6$ redshift dependence, in sources characterized by Eddington or super-Eddington accretion flows \citep{kawanaka2021}. In these sources the radiation-driven wind can act as a low-temperature,  optically thick corona where the optical depth is larger for higher mass outflow rate winds. This give rise to progressively softer (steeper) spectra. If this is indeed the scenario, then this could be an indication that these sources are accreting at super critical rates, much larger than what $\ledd$ may imply.

\subsection{Implications for the $z>6$ AGN population and their detectability in the X-rays}

Our result represents one of the most significant difference reported so far between QSOs at EoR and those at lower redshifts. Given the sample selection, in principle this result should be only valid for the QSO population whose SMBHs underwent a fast mass growth history. This includes also the recently discovered $z\approx8-10$ JWST AGN \citep{larson2023,maiolino2023,bogdan2023} which would require $\mseed=10^3-10^4 \msun$ (in Fig.~\ref{hyperion_sample}, left) and hence are expected to have experienced fast mass growth.
The mild indication (at the $2\sigma$ level) of an increasing $\Gamma$ with increasing $\mseed$, if confirmed, could imply a flattening of $\Gamma$ for QSOs requiring a less extreme $\mseed$/mass growth. This would reconcile the X-ray properties of the less extreme \hyp\ QSOs with those reported for normal lower-z QSOs. However the real fraction of sources with low $\mseed$ and the presence of a relation between $\Gamma$ and $\mseed$ are still an open issue.  
Hence, we cannot exclude that our results may apply to the entire $z>6$ QSO or AGN population. If this is the case, this may have an important impact on the source detectability in future X-ray surveys and on our capabilities to study and understand nuclear accretion mechanisms at EoR. 

 Indeed, the predicted 0.5-2~keV and 2-10~keV fluxes for sources with a given  0.5-10~keV luminosity are a factor of $\sim1.9$ and $\sim4.1$ fainter assuming a power-law with $\Gamma=2.4$ instead of a canonical $\Gamma=1.9$ value. Alternatively, assuming a power-law with a $E_{cut}$ at 20~keV and $\Gamma=1.9$ at the average redshift of the \hyp\ sample, $z=6.7$, these factors change to $\sim1.3$ and $\sim4.2$, respectively. 
 In this case, at higher redshifts (e.g. $z=8$) the detection would be even harder as the factors would increase to $\sim1.4$ and $\sim5.3$, respectively. 
 
This issue must be considered when defining the sensitivity capabilities and the design of next-generation X-ray observatories (e.g. ATHENA, Lynx) especially aiming at the highest-redshift Universe that we are currently probing with JWST. This is a fundamental step to extend our current understanding of the yet undiscovered AGN population of which the QSOs we are currently studying may represent the tip of the iceberg. Indeed, recent JWST observations suggest that previously known high-redshift star-forming galaxies may harbor an AGN in their nucleus \citep{cameron2023,maiolino2023,ubler2023}.  The disclosure of these previously "hidden" AGN may have a role in explaining the high UV luminosity density of luminous galaxies (Trinca et al. in prep.) observed at $z\gtrsim8$ \citep{donnan2023,harikane2023,bouwens2023}.

\section{Conclusions and future prospects}\label{conclusions}
In this paper, we have presented the \hyp\ sample of  QSOs at the Epoch of Reionization selected for their fast SMBH growth history. Indeed, \hyp\ QSOs are powered by SMBHs that would descend from seeds  of $\mseed>10^3\rm~\msun$ at $z=20$, if accreting continuously at the Eddington rate. 
The sample consists of 18 QSOs at redshifts $z\approx6-7.5$ (mean $z\sim6.7$) with average luminosity $\lbol\approx10^{47.3}~\rm \lumcgs$ and $\mbh\approx10^9-10^{10}~\rm \msun$.
 \hyp\  builds on a 2.4~Ms \xmm\ Multi-Year Heritage Program (three years) designed to accurately characterize, for the first time on a statistically sound sample, the X-ray nuclear properties of QSOs at the Epoch of Reionization.
In this paper we report on the spectral analysis of the first year of \xmm\ observations  of \hyp. We have analyzed \xmm\ observations of twelve \hyp\ QSOs for a total exposure of $\sim1$~Ms. New \hyph\ observations for ten  sources have been presented in this paper. These include the first X-ray detection/spectral analysis ever reported for two sources (J083.8 and J029). All the QSOs have been detected with the exception of two sources (J231.6 and J011). 

Our main findings are summarized as follows:
\begin{itemize}
    \item the X-ray spectral analysis on individual sources using simple power-law models on spectra with 50-140 net-counts (pn+MOS, 0.3-7~keV) resulted in a wide range of $\Gamma\approx1.9-3$, with the large majority of sources (80\%) exhibiting steep $\Gamma\gtrsim2.1$;
    \item a power-law joint spectral analysis of the 10 detected sources resulted in an average $\Gamma\approx2.4\pm0.1$. Moderate absorption (i.e. $\sim10^{21}-10^{22}~\nhi$) is not required. The steep $\Gamma$ value rules out for the first time (at the $\sim4~\sigma$ level) a canonical $\Gamma=2$ and the average $\Gamma$ reported in $z<6$ QSOs of similar luminosity or $\ledd$. This implies that the steepness of the X-ray spectrum in \hyp\ QSOs is an evolutionary signature of the \hyp\ QSOs regardless of the QSO radiative output and SMBH accretion rate;
    \item a joint spectral analysis with a $\Gamma=1.9$ power-law and a high energy cutoff model resulted into a very low-energy $E_{cut}\sim20$~keV. This value, if confirmed with future \hyph\ data, is unreported at such high luminosities and redshifts;
    \item The X-ray bolometric correction is in line with the trend reported for the bulk of AGN at high luminosity regimes. However, we find that the optical-to-X-ray spectral index $\aox$ (and $\Delta \aox$) is slightly higher than the $\aox$~vs.~$\luvmon$ relations reported for large AGN samples. This is a consequence of the \hyp\ QSOs steep X-ray spectral slopes. 
\end{itemize}

    We interpret the steep spectral slopes as an indication of cool coronae originated either by (i) the interaction with the soft radiation field of the accretion disc or (ii) the peculiar properties of the X-ray corona itself. 
    In the first case the disc is supposed to be highly accreting or truncated in the inner regions possibly by nuclear winds with high mass outflow rate. Alternatively, enhanced cooling may be due to multiple scattering in high optical depth coronae, or to highly energetic non-thermal electrons cooling and interacting with thermal electrons in hybrid coronae models. 
    We think that the inner disc truncation scenario by disc winds with a high mass flux offers at the moment, unlike other scenarios, a robust explanation of the $\Gamma$ redshift evolution as it relies on redshift-dependent results found for nuclear winds.

The \hyph\ program presented here constitute a remarkable leap forward in the nuclear characterization of QSOs at EoR. 
More \xmm\ data are coming. They will strengthen our findings and extend them to the entire \hyp\ sample. This will allow us to assess the rate at which $\Gamma$ steepens as a function of redshift or the mass growth history (adopting $\mseed$ as a proxy). Broad-band UV/X-ray physical models for the continuum from the accretion disc/corona system will be applied to these data jointly with quasi-simultaneous UV rest-frame data which we are progressively collecting during the three years of the \hyph\ program. 
Our data will enable us to shed light on the coupling between the X-ray and the broad-line properties. Furthermore together with sub-mm/mm data we will be able to investigate the impact of the QSO X-ray radiation field affecting the excitation of the molecular medium, that can be constrained by targeting high-J transitions of the CO molecule rotational ladder. 

If the nuclear properties reported here for the \hyp\ QSOs are confirmed to be typical of the whole QSO population at EoR, the distinctive steep spectral slopes obtained from our analysis will have an important impact on the source detectability in future X-ray surveys. In particular, the soft and hard band fluxes for sources of a given luminosity will be $\sim2$ and $\sim4$ times fainter then expected for a power-law with canonical flatter spectral index. The hard band fluxes will be even fainter at higher redshifts if the true spectral model consist in standard $\Gamma=1.9$ power-law with low-energy ($\sim20$~keV) cutoff. 
Accounting for this will have an impact on the design and the capabilities of the future X-ray flagship observatories aiming at probing nuclear accretion in the early Universe.

\begin{acknowledgement}
We thank the anonymous referee for the constructive comments which helped clarify several aspects of the paper.
  The analysis and results presented in this work are based on observations obtained with \xmm\, an ESA science mission with instruments and contributions directly funded by ESA Member States and NASA. 
We thank the \xmm\ Science Operation Centre and Norbert Schartel for their help, prompt support and advise for the scheduling and optimisation of the \hyph\ program.

The authors acknowledge financial support from the Bando Ricerca Fondamentale INAF 2022 Large Grant "Toward an holistic view of the Titans: multi-band observations of z>6 QSOs powered by greedy supermassive black-holes". LZ and FT acknowledge support from INAF “Progetti di Ricerca di Rilevante Interesse Nazionale” (PRIN), Bando 2019 “Piercing through the clouds: a multiwavelength study of obscured accretion in nearby supermassive black holes”. GM was partly supported by grant n. PID2020-115325GB-C31 funded by MCIN/AEI/10.13039/501100011033. CD acknowledges support from the STFC consolidated grant ST/T000244/1. We thank Stefano Bianchi, Ciro Pinto and Luigi Stella for useful discussions. 
\end{acknowledgement}

\bibliographystyle{aa}
\bibliography{hyperion}

\begin{appendix}

\section{\xmm\ EPIC detectors images of the \hyph\ QSOs in the 0.5-2~keV}

We present here pn, MOS1 and MOS2 0.5-2~keV images of the observations reported in Table~\ref{obsjournal}. For each detector we report source and background region files adopted for photometry and spectral extraction. The figure for exposure $\rm J1342\_1$, i.e. the first observation of J1342 is reported in Fig.~\ref{hyperion_subsample}.
\begin{figure*}[!t]
   \begin{center}
    \includegraphics[width=0.85\textwidth, angle=0]{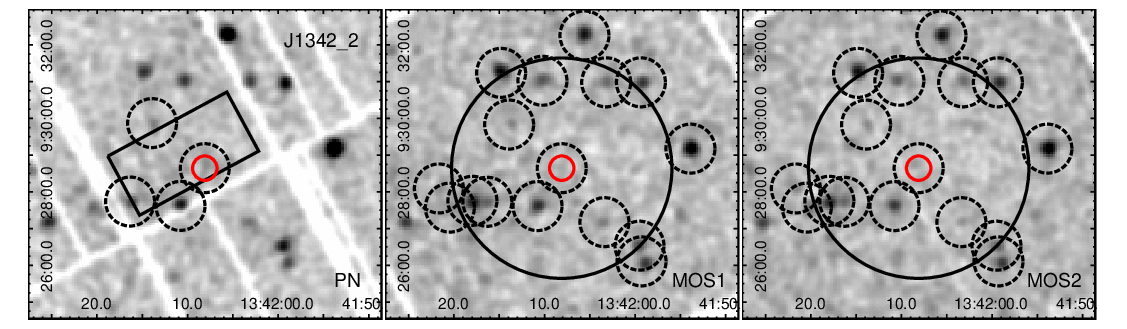}
    \includegraphics[width=0.85\textwidth, angle=0]{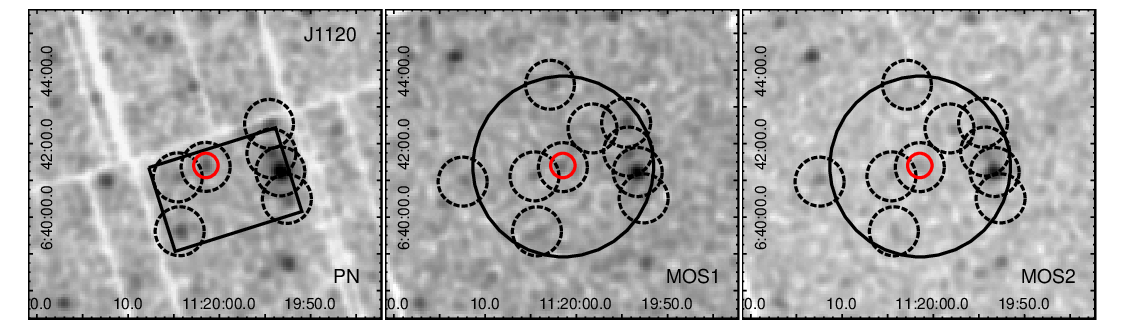}
    \includegraphics[width=0.85\textwidth, angle=0]{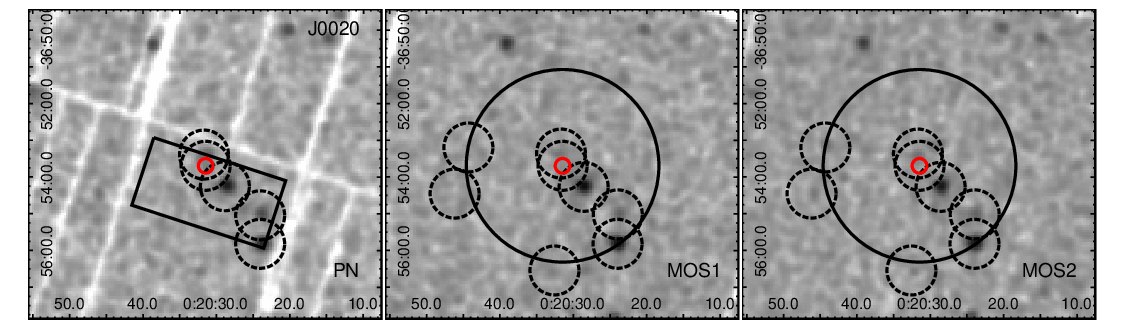}
    \includegraphics[width=0.85\textwidth, angle=0]{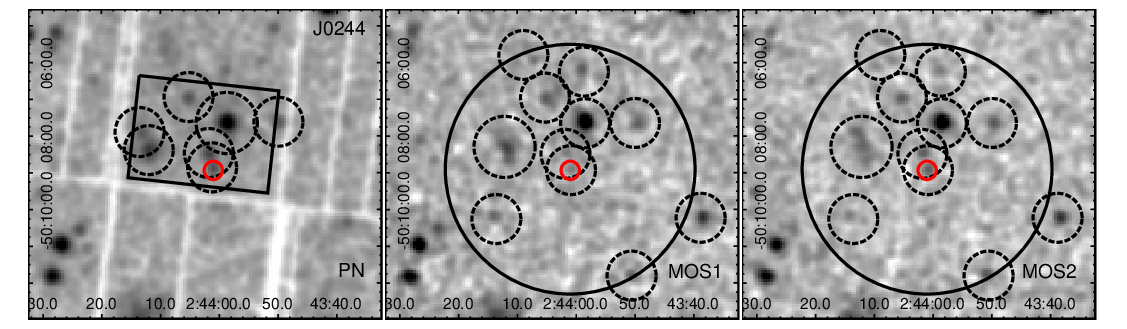}
    \includegraphics[width=0.85\textwidth, angle=0]{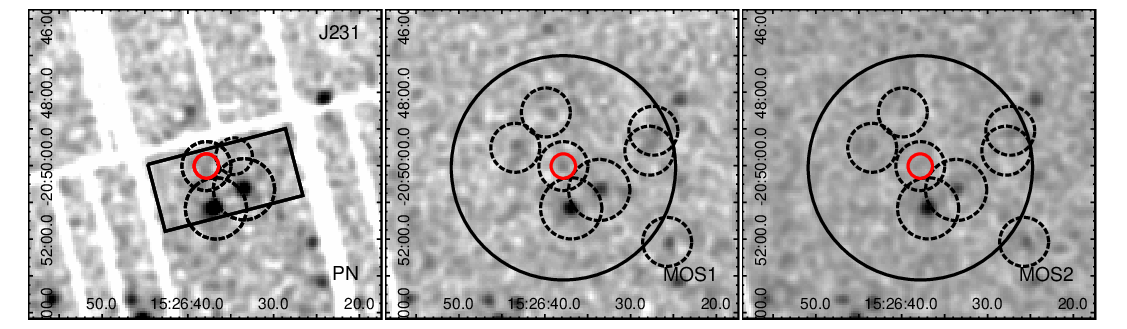}
   \end{center} 
       \label{hyperion_subsample_others}
\caption{EPIC 0.5-2~keV pn, MOS1 and MOS2 camera images for sources of the \hyph\ program presented in this work (J1342\_1 is reported in Fig.~\ref{hyperion_subsample}). All the images are smoothed by a Gaussian kernel of 3~pixel radius for better visualization. Source and background counts/spectral extraction regions are reported in red and black, respectively. Dashed circular regions indicates areas excluded by the background extraction.}
\end{figure*}
\setcounter{figure}{0}
\begin{figure*}[t!]
   \begin{center}
    \includegraphics[width=0.85\textwidth, angle=0]{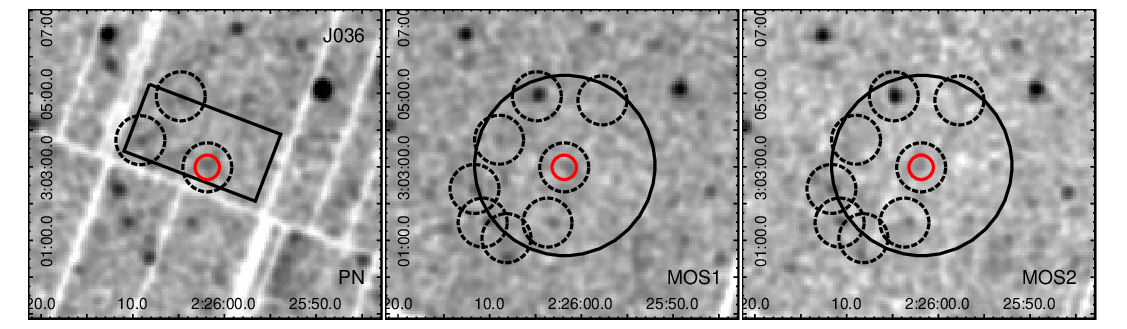}
    \includegraphics[width=0.85\textwidth, angle=0]{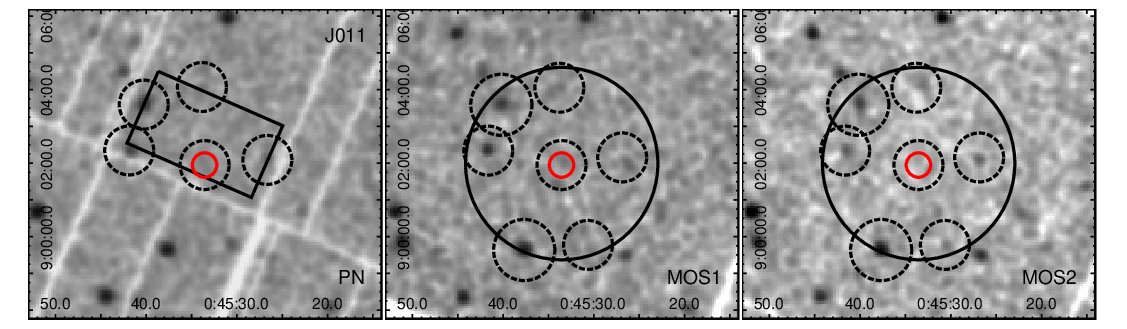}
    \includegraphics[width=0.85\textwidth, angle=0]{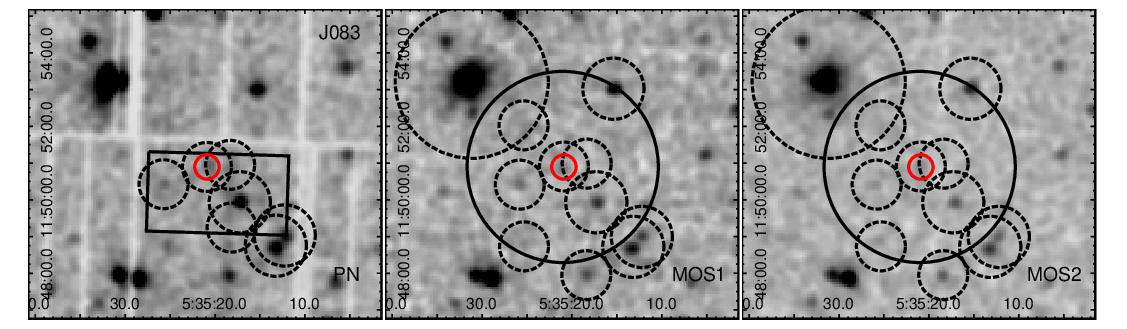}
    \includegraphics[width=0.85\textwidth, angle=0]{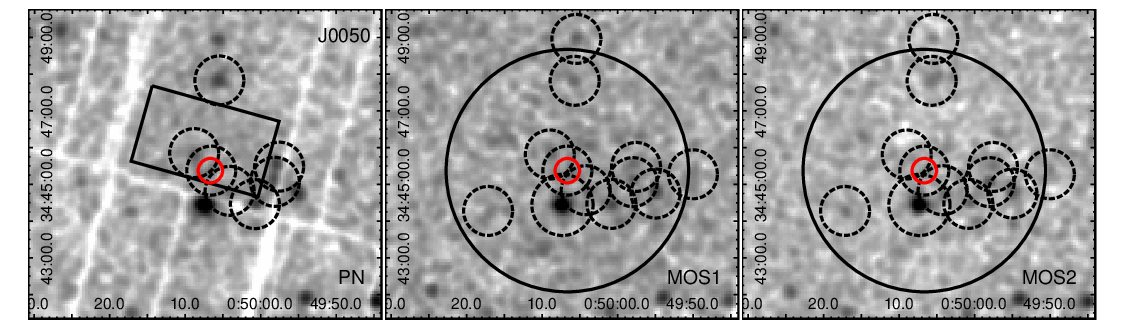}
    \includegraphics[width=0.85\textwidth, angle=0]{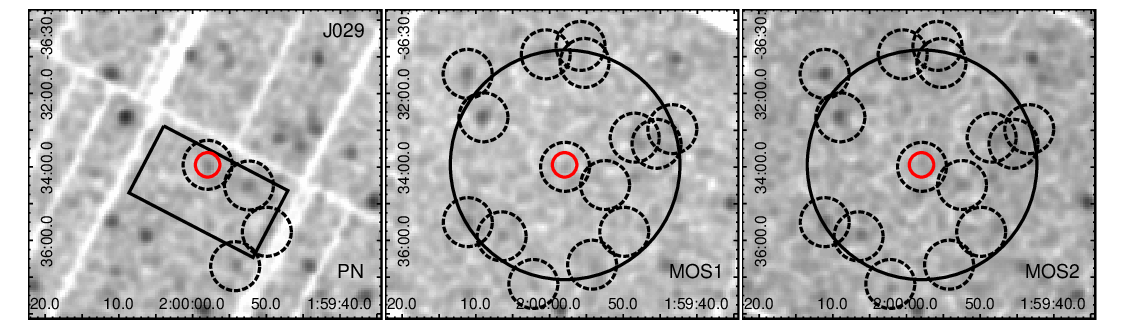}
   \end{center} 
\caption{continued}
\end{figure*}

\setcounter{figure}{0}
\begin{figure*}[t!]
   \begin{center}
    \includegraphics[width=0.85\textwidth, angle=0]{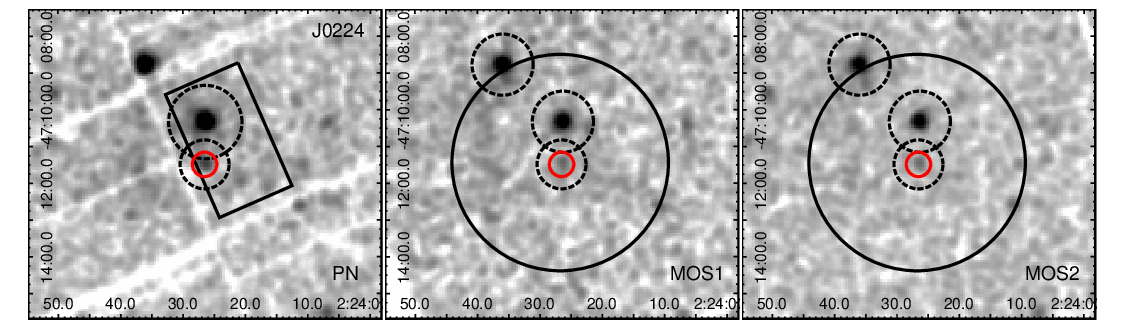}
    \includegraphics[width=0.85\textwidth, angle=0]{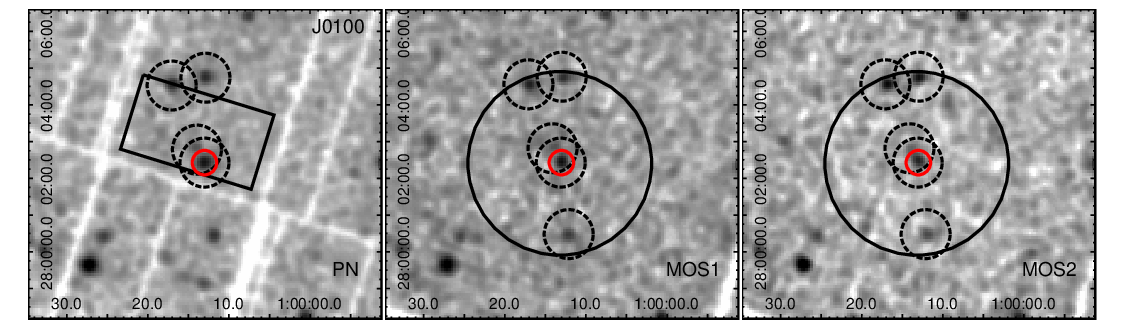}
   \end{center} 
\caption{continued}
\end{figure*}


\section{Optimal spectral binning for low counts spectra}\label{optimalbin}
In order to evaluate the optimal binning for our spectra we simulated spectra for steep and flat $\Gamma$ and evaluated the accuracy (i.e. the difference between input and best-fit simulated value in units of the input value) in recovering  the input $\Gamma$ and $\lumh$ values. We take as a reference, the spectra in our sample with 60-70~net-counts (pn+MOS, 0.3-10~keV). Specifically, for the steep and flat $\Gamma$ we adopted  the input best-fit values reported in Table~\ref{analysis} for J029 and J0050, respectively. 
Hence we simulated 10000 set of spectra for each $\Gamma$ case evaluating the following binning schemes: unbinned, binned at minimum  1,3,5 and 10 counts and the optimal sampling from \citet{kaastra2016}, hereafter called KB. For each binning scheme, we also evaluated the energy range dependence in the following intervals 0.3-2~keV, 0.3-5~keV, 0.3-7~keV and 0.3-10~keV.
\begin{figure*}[b!]
   \begin{center}
    \includegraphics[width=0.4\textwidth, angle=0]{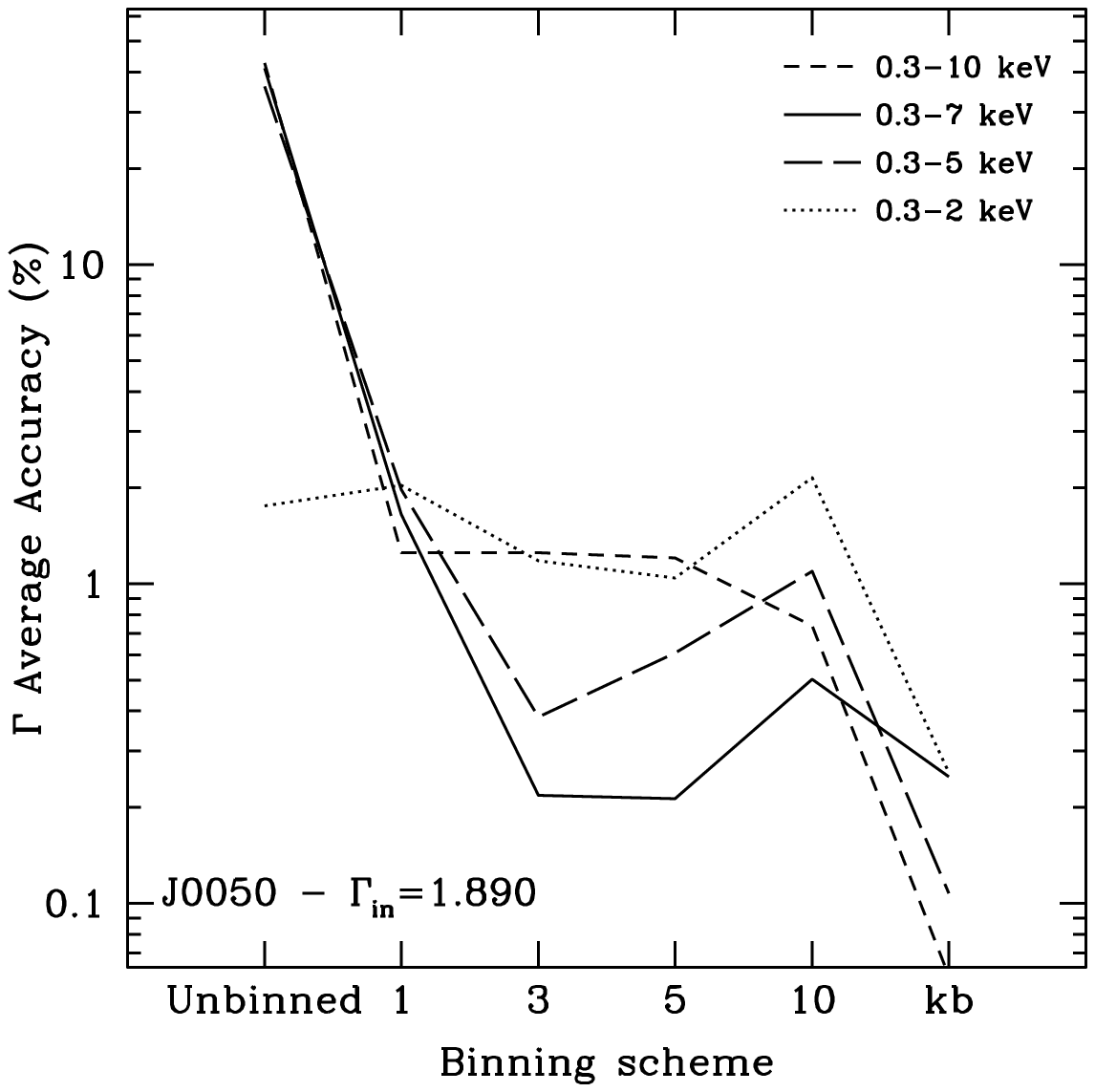}
    \includegraphics[width=0.4\textwidth, angle=0]{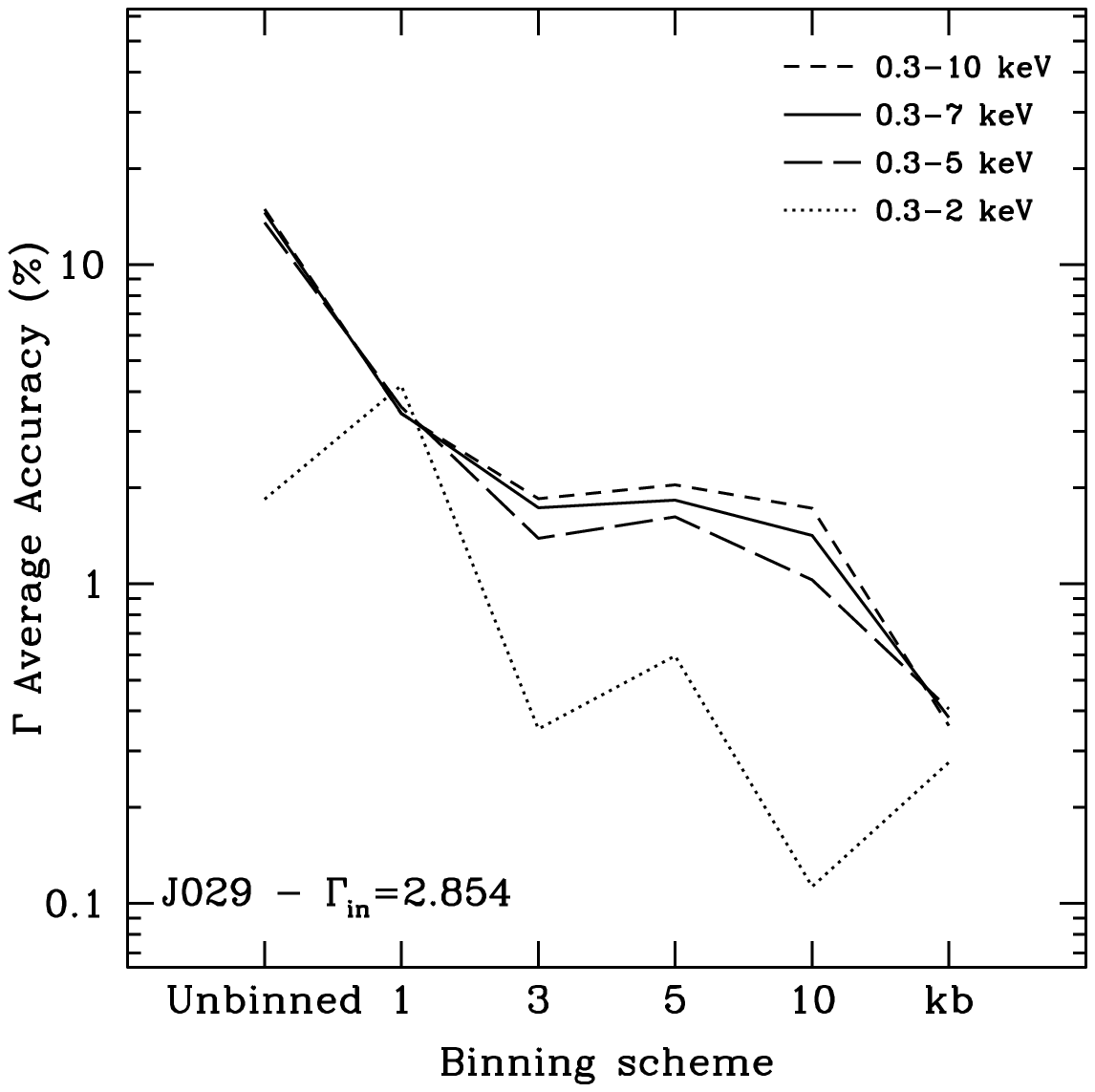}
    \includegraphics[width=0.4\textwidth, angle=0]{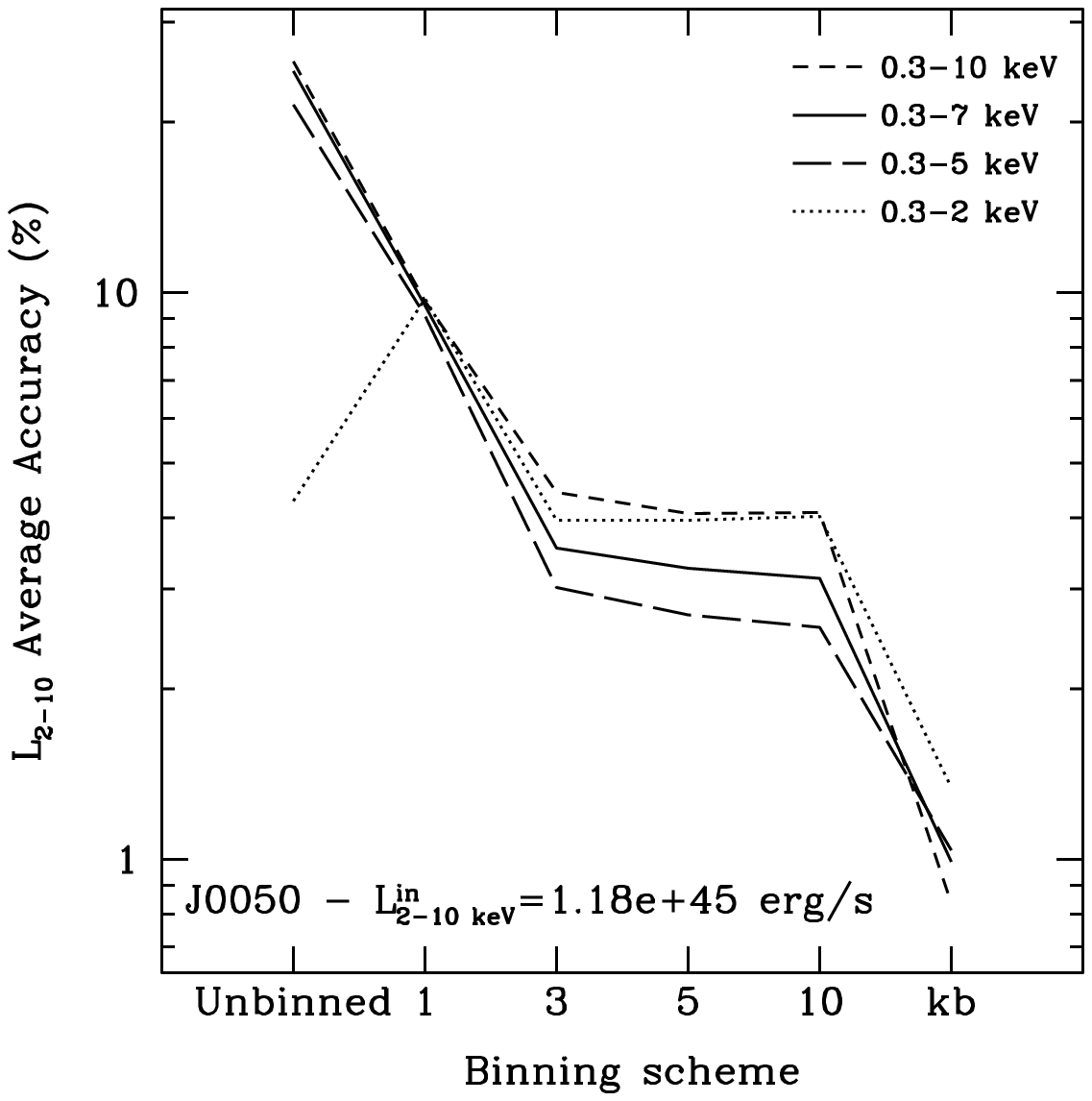}
    \includegraphics[width=0.4\textwidth, angle=0]{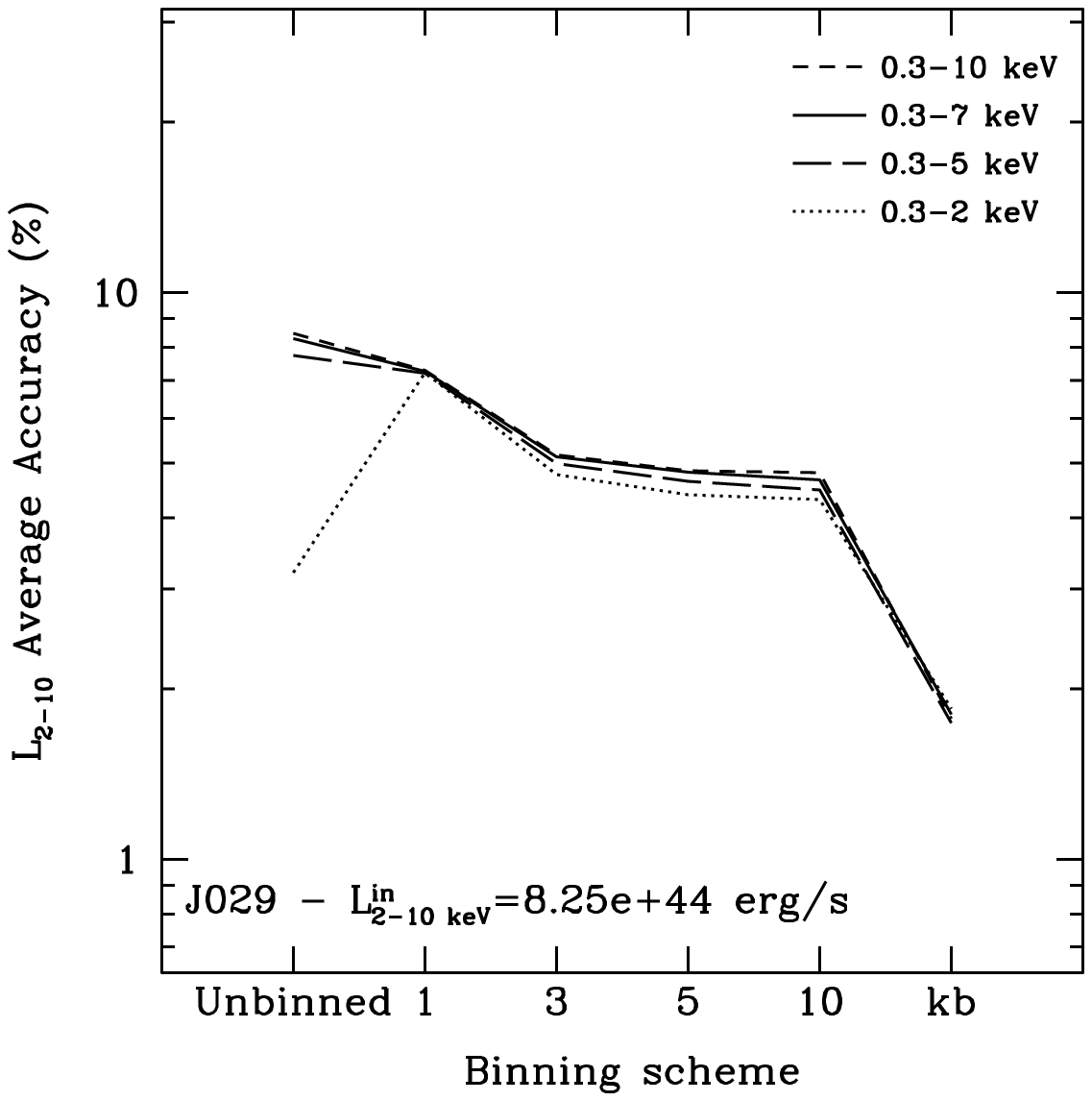}
   \end{center} 
\caption{Accuracy, as a function of binning scheme and energy range, in recovering the input $\Gamma$ and $\lumh$ through a set of 10000 spectral simulations. }
       \label{binning_simulations}
\end{figure*}
In general, we found that the binning scheme plays a negligible role in the accuracy of our results, especially compared to the size of our statistical errors (i.e. the scatter of the distribution of the best-fit values), which are always almost a factor of few up to 1-2 orders of magnitude larger at all energy intervals probed. Hence possible systematics in the fitting process are in general compensated by the larger statistical uncertainties. Fig.~\ref{binning_simulations} show the accuracy in recovering the average $\Gamma$ and $\lumh$ as a function of binning and energy range.
In general, unbinned results are very inaccurate especially in energy intervals including background-dominated upper energy bounds (i.e. $> 2~\rm keV$) and can bias the recovered values by more than 10\%. 
On average, going to larger bins improves the accuracy of the parameter estimation to sub-percent or percent level for flat or steep input $\Gamma$, respectively, and to few percent level in case of $\lumh$. 
The KB binning scheme consistently recovers at least a factor of 2-3 more accurate results at all energies  regardless of the input $\Gamma$ (showing larger accuracies for flat $\Gamma$).

\section{Comparison with previous analysis of the \hyph\ detected source} \label{cfrprevious}
In this section we will compare our best-fit $\Gamma$ and $\lumh$ reported in Table~\ref{analysis} with previous analysis carried on the same sources with already archived observations.

J1342. A 45~ks \chandra\ observation of J1342 has been analyzed by \citet[][]{banados2018b} and \citet[][]{vito2019}. The source was detected with $\sim$14 net-counts (0.5-7 keV). They attempted a basic spectral analysis with a power-law model with Galactic absorption with similar results. \citet[][]{vito2019} found $\Gamma\approx1.97^{+1.16}_{-0.92}$ and $\lumh=14.95^{+11.51}_{-7.60}\times10^{44}~\lumcgs$ (errors at 90\% level). These values are consistent at $\lesssim1\rm~\sigma$ level.

J1120. A $\sim340$~ks \xmm\ observation divided in three exposures has been analyzed by several authors \citep{page2014,moretti2014,nanni2017,vito2019}. The observation resulted heavily contaminated by background flares ($\sim50$\% in pn). The last analysis by \citet{vito2019} obtained $\Gamma=2.08^{+0.74}_{-0.64}$ and $\lumh=6.56^{+3.59}_{-3.27}\times10^{44}~\lumcgs$ (errors at 90\% level). The $\Gamma$ is consistent a $\sim1\sigma$ level with our value. The luminosity $\lumh$ is inconsistent at $\sim4.5\sigma$ level. Hence the source appears to have increased its luminosity by a factor of $\sim4$.

J0020. A 25~ks \xmm\ observation was analyzed by \citet{pons2020}. According to the authors the source resulted undetected with a $\lumh<4.76\times10^{45}~\rm \lumcgs$ upper limit. Their estimate is consistent with our luminosity value.

J0244. A  17~ks \xmm\ observation was analyzed by \citet{pons2020}. According to the authors the source resulted undetected with a $\lumh<4.37\times10^{45}~\rm \lumcgs$ upper limit. Their estimate is consistent with our luminosity value.

J036.5. A 25~ks \chandra\ observation was analyze by \citet{vito2019}. The source was detected with $5.5$ net-counts. They attempted a spectral analysis obtaining a $\Gamma\approx2.1^{+2.2}_{-1.5}$ and $\lumh < 20.53\times10^{44}\rm \lumcgs$. A $\sim17$~ks \xmm\ observation was also analyzed by \citet{pons2020}. They did not detect the source and only obtained a very high upper limit on the luminosity of $\lumh<17.62\times10^{45}\rm~\lumcgs$. All measurements are consistent with our best-fit values.

J0050. A 34~ks \chandra\ observation was analyzed by \citet{vito2019}. The source was detected with 7.4~net-counts. They attempted a spectral analysis obtaining a $\Gamma\approx2.1^{+2.0}_{-1.2}$ and $\lumh = 8.2^{+8.8}_{-5.0}\times10^{44}\rm \lumcgs$. Their values are consistent with ours at $<1\sigma$ level.

J0224. For this source only a 26~ks \xmm\ is available in the archive. This is the observation we have analysed in this work. A previous analysis of this observation was carried out by \citet{pons2020}. They obtained $\Gamma=1.82^{+0.29}_{-0.27}$ and $\lumh=2.92\pm0.43\times10^{45}\rm~\lumcgs$. These values are consistent at $<1.2\sigma$ level with ours.

J0100. For this source a 15~ks \chandra\ observation and a $\sim65$~ks \xmm\ observation are archived. We have analysed the longer \xmm\ observation which provides a factor $>20$ more net-counts than the \chandra\ one. The \xmm\ observation was analyzed by \citet{ai2017} and by \citet{vito2019}. The latter obtained $\Gamma=2.52^{+0.23}_{-0.22}$ and $\lumh=67.55^{+9.63}_{-8.93}\times10^{44}\rm~\lumcgs$ (errors at 90\% level). These values are consistent with our analysis. Their $\Gamma$ is consistent at $<1\sigma$ level while the luminosity show consistency at  $\sim1.5\sigma$ level.

\end{appendix}

\end{document}